\documentclass[twocolumn,twocolappendix]{aastex701}

\usepackage{amsmath}
\usepackage{amssymb} \usepackage{bm}
\usepackage{graphicx}
\graphicspath{{./}}
\newcommand{\Cov}{\operatorname{Cov}}

\begin{document}
{\color{black}
\title{On the Spectrum of MHD Turbulence and Alignment}

\correspondingauthor{Amir Jafari}
\author{Amir Jafari}
\affiliation{Ashdown Lodge, Notting Hill, London, United Kingdom}
\email{elenceq@jhu.edu}

\begin{abstract}
The original dynamic-alignment phenomenology attributed the $k_\perp^{-3/2}$ spectrum, observed in some simulations, to a scale-dependent decrease of the characteristic angle between perpendicular velocity and magnetic increments. However, numerical studies find scale dependence only for a weighted mean angle, such as the amplitude-weighted angle between perpendicular Els\"asser increments, $\langle s_r\rangle_A=\langle A_rs_r\rangle/\langle A_r\rangle$, where $A_r=|\delta_r\boldsymbol z_\perp^+||\delta_r\boldsymbol z_\perp^-|$ and $s_r=\sin\theta_r$. Higher-amplitude increments correlate with smaller unweighted angles thus $\langle s_r\rangle_A$ can change with scale through changes in the joint amplitude--angle statistics even when the unweighted mean changes little. We derive an exact decomposition of the change in $\langle s_r\rangle_A$ between two scales into an angular term $C_\theta$, an amplitude term $C_A$, and a joint term $C_\times$ and use the joint distribution of $(A_r,s_r)$ to measure angular change at fixed initial angle as a function of amplitude. It follows that the spectral condition is independent of the weighted-angle condition thus a $k_\perp^{-3/2}$ spectrum can coexist with either a scale dependent or scale independent $\langle s_r\rangle_A$. We test these results in full-MHD and reduced-MHD simulations and NASA Wind measurements. The fixed-scale covariance is negative in all three and in every interval where the weighted angle decreases, both $C_\theta$ and $C_A$ are non-negative while $C_\times$ is negative and larger in magnitude. At fixed initial angle, higher-amplitude increments undergo smaller angular changes but geometry alone can produce this ordering. Using these results, we argue that amplitude-angle anticorrelation may arise from forcing in the injection range. 
\end{abstract}

\section{Introduction}
\label{sec:introduction}

The energy spectrum is a crucial quantity in both hydrodynamic and magnetohydrodynamic (MHD) turbulence because it shows how energy is distributed among scales. Let $E(k_\perp)$ denote the fixed-plane one-dimensional perpendicular spectrum;
$$
 \frac{1}{2}\left\langle|\boldsymbol u_\perp|^2+|\boldsymbol B_\perp|^2\right\rangle=
 \frac{1}{4}\left\langle|\boldsymbol z_\perp^+|^2+|\boldsymbol z_\perp^-|^2\right\rangle
 =\int_0^\infty E(k_\perp)\,d k_\perp,
$$
with Els\"asser variables $\boldsymbol z^\pm=\boldsymbol u\pm\boldsymbol B$, so $\boldsymbol z_\perp^\pm=\boldsymbol u_\perp\pm\boldsymbol B_\perp$, and $\boldsymbol B$ in Alfv\'en-speed units. The distinction from the local-field structure functions used in the full-MHD analysis is discussed in Appendix~\ref{app.spectra}. The physical mechanisms responsible for the spectrum and exact conditions under which different scalings may arise remain open questions, and several phenomenological descriptions have been proposed.

 Under the critical balance assumption and a scale-independent energy flux \citep{GoldreichSridhar1995}, Els\"asser increments scale as $|\delta_r\boldsymbol z_\perp^\pm|\propto r^{1/3}$, which leads to the $k_\perp^{-5/3}$ scaling (Appendix~\ref{app.spectra}). The exponent $1/3$ is the Kolmogorov scaling \citep{Kolmogorov1941}, which also coincides with the critical H\"older exponent for anomalous dissipation in Onsager's theory and its extension to ideal MHD \citep{Onsager1949,CaflischKlapperSteele1997}. Some numerical simulations report perpendicular spectra close to this Kolmogorov-type spectrum \citep{ChoLazarianVishniac2002,ChoLazarian2003}; \cite{BiskampMuller2000} found the same exponent for an isotropic shell spectrum. However, a $k_\perp^{-3/2}$ spectrum was also reported in some other simulations~\citep{MaronGoldreich2001,MullerGrappin2005}. The Iroshnikov--Kraichnan phenomenology also predicts a $k^{-3/2}$ spectrum, but assumes three-dimensional isotropy and progressively weak interactions and therefore does not describe this anisotropic strong-turbulence regime~\citep{Iroshinkov1964,Kraichnan1965}.

In an attempt to explain this scaling in an anisotropic cascade,~\cite{Boldyrev2005,Boldyrev2006} proposed that nonlinear interactions between counter-propagating Els\"asser fluctuations are weakened if the velocity and magnetic field fluctuations tend to align progressively toward smaller scales in the inertial range---a scale-dependent form of dynamic alignment, distinct from the older use of the term for the temporal evolution of decaying MHD toward an Alfv\'enic state \citep{Mason2006}. Specifically, for a strong guide field, the nonlinear interaction in this picture is weakened by a factor proportional to the sine of the characteristic angle between velocity and magnetic increments in the plane perpendicular to the local magnetic field, $\delta_r\boldsymbol u_\perp$ and $\delta_r\boldsymbol B_\perp$, and this angle decreases toward smaller separations as $r^{1/4}$, leading to the shallower $k_\perp^{-3/2}$ spectrum. 

The alignment argument assumes that the  angle is of the order of \(\theta_\lambda\sim\delta B_\lambda/B_0\), where $\delta B_\lambda$ is the magnetic fluctuation at perpendicular scale $\lambda$ and $B_0$ is the guide-field strength. This is at odds with the reduced MHD (RMHD) symmetry \citep{Beresnyak2011}, while simulations show little dependence of the alignment on this ratio \citep{BeresnyakLazarian2009}. Later work reported that the measured alignment saturates and does not converge to the predicted scaling at higher resolution \citep{Beresnyak2011,Beresnyak2012}. However, \cite{PerezEtAl2012} attributed this flattening to an under-resolved dissipation range. 

It is crucial to recognize that the scale dependence of the mean alignment angle can depend on the choice of weighting, which is precisely the focus of the present work. The original proposal \citep{Boldyrev2005,Boldyrev2006} invoked a characteristic angle between velocity and magnetic fluctuations, while the nonlinear interaction was expressed through the angle between counter-propagating Els\"asser fluctuations. These angles are related but not identical, and neither was defined as a precise statistical measure. A numerical test requires a specific definition, for example an ensemble mean that is either unweighted or weighted by some function. The angle itself can be independent of scale, yet its weighted mean can become scale dependent simply because the correlation between the weight and angle changes with scale.

Different simulations have used different measures, including unweighted averages or rms measures of the sine of the angle, and more often amplitude-weighted measures \citep{BeresnyakLazarian2006,BeresnyakLazarian2009,Mason2006}. A scaling close to $r^{1/4}$ was reported for a strong guide field using an amplitude-weighted measure \citep{Mason2006}, whereas the unweighted measures showed weaker scale dependence \citep{BeresnyakLazarian2006,BeresnyakLazarian2009}. The ratio
\begin{equation*}
\frac{\left\langle|\delta_r\boldsymbol u_\perp\times\delta_r\boldsymbol B_\perp|\right\rangle}
 {\left\langle|\delta_r\boldsymbol u_\perp||\delta_r\boldsymbol B_\perp|\right\rangle}
=
\frac{\left\langle|\delta_r\boldsymbol u_\perp||\delta_r\boldsymbol B_\perp|\sin\alpha\right\rangle}
 {\left\langle|\delta_r\boldsymbol u_\perp||\delta_r\boldsymbol B_\perp|\right\rangle}
\end{equation*}
where $\alpha$ is the angle between $\delta_r\boldsymbol u_\perp$ and $\delta_r\boldsymbol B_\perp$, is the ensemble mean of $\sin\alpha$ weighted by
$|\delta_r\boldsymbol u_\perp||\delta_r\boldsymbol B_\perp|$
\citep{Mason2006,BeresnyakLazarian2009}, therefore larger amplitudes contribute more strongly. While some simulations
found a scaling close to $r^{1/4}$ for this weighted mean
\citep{Mason2006,PerezEtAl2012}, others questioned whether this
scaling is asymptotic and even whether the weighted mean determines
the energy spectrum
\citep{BeresnyakLazarian2009,Beresnyak2012}.
\cite{BeresnyakLazarian2006} found weak scale dependence of the
rms geometrical alignment of the Els\"asser fluctuations, while
their fourth-order measures revealed scale-dependent polarization
intermittency and interaction weakening. \cite{BeresnyakLazarian2009} also reported a sub-$1/4$ scaling for the Els\"asser-weighted measure used below. Other amplitude weightings and moment orders have also been studied \citep{ChandranSchekochihinMallet2015,MalletSchekochihinChandran2015,MalletEtAl2016}. 

Despite the ongoing debate about the spectrum of MHD turbulence and the angle between fluctuations, several results seem to be well established. For a review, see \cite{Schekochihin2022}. At fixed separation $r:=|\boldsymbol r|$ in the inertial range, higher-amplitude fluctuations correlate with smaller mutual (unweighted) angles $\theta_r$ and greater local anisotropy \citep{BeresnyakLazarian2006,MalletSchekochihinChandran2015,MalletEtAl2016}, while intermittent reduced-MHD theories relate the same quantities \citep{ChandranSchekochihinMallet2015,MalletSchekochihin2017}. Unweighted and amplitude-weighted angle measures can have different scale dependencies, and different amplitude weightings and moment orders can give different measured scalings \citep{BeresnyakLazarian2006,BeresnyakLazarian2009,Beresnyak2012,ChandranSchekochihinMallet2015,MalletEtAl2016,MalletSchekochihin2017}. Solar-wind measurements have reported alignment toward smaller scales over part of the inertial range \citep{Podesta2009}, while more recent Wind measurements find increasing misalignment within the inertial range and show that the scaling depends on whether the Els\"asser or velocity--magnetic angle is used \citep{SioulasEtAl2024}.

Let us adopt the notation
$$
A_r:=\left|\delta_r\boldsymbol z^{+}_{\perp}\right|
\left|\delta_r\boldsymbol z^{-}_{\perp}\right|,\,
s_r:=\sin\theta_r,\,
\langle s^q_r\rangle_{A^p}:=
\frac{\langle A^p_r s^q_r\rangle}{\langle {A^p_r}\rangle},$$
where $0\leq\theta_r\leq\frac{\pi}{2}$ is the folded angle
between Els\"asser increments $\delta_r\boldsymbol z^+_\perp$ and
$\delta_r\boldsymbol z^-_\perp$, in the plane perpendicular
to the local magnetic field $\boldsymbol B_L$, so that alignment
and anti-alignment are treated alike.  Simple algebra yields
\begin{align}
 \left\langle s_r^q\right\rangle_{A^p}
 &=\left\langle s_r^q\right\rangle
 +\frac{\Cov(A_r^p,s_r^q)}{\left\langle A_r^p\right\rangle}.
 \label{eq:weightedhierarchy}
\end{align}
Thus $p$ controls the amplitude weighting and $q$ the angular moment. We retain $p\geq0$ and $q>0$ in the formal decomposition but use $p=q=1$ in its numerical evaluation and several values of $p$ only to examine amplitude weighting.

Equation~\eqref{eq:weightedhierarchy} is our starting point. It shows that a negative covariance $\Cov(A_r,s_r)$ automatically places $\langle s_r\rangle_A$ below $\langle s_r\rangle$ at any given scale. It also shows that a weighted mean can depend on scale while the unweighted mean remains independent of scale. Our numerical results indicate a nearly scale-independent unweighted mean $\langle s_r\rangle$, shown in Figures~\ref{figure;higherMoments} and~\ref{fig_app_weighted_covariance}, in agreement with other simulations~\citep{BeresnyakLazarian2006,BeresnyakLazarian2009,Beresnyak2011}. Thus a scale-dependent $\langle s_r\rangle_A$ does not by itself imply a scale-dependent $\langle s_r\rangle$; it may instead reflect the scale dependence of $\Cov(A_r,s_r)/\langle A_r\rangle$.

But how can the weighted mean $\langle s_r\rangle_A$ become scale-dependent? With a scale-independent $\langle s_r\rangle$, Equation~\eqref{eq:weightedhierarchy} forces us to turn to the amplitude $A_r$ and its correlation with $s_r$. It is then natural to start with the Hölder scaling $|\delta_r\boldsymbol z_\perp^\pm|\sim r^{h^\pm}$ which enters $A^p_r=|\delta_r\boldsymbol z_\perp^+|^p|\delta_r\boldsymbol z_\perp^-|^p  \sim r^{p(h^++h^-)}$. For constant Hölder exponents $h^\pm$, the explicit factor $r^{p(h^++h^-)}$ cancels between the numerator and denominator in the weighted mean $\langle s^q_r\rangle_{A^p}=
{\langle A^p_r s^q_r\rangle}/{\langle {A^p_r}\rangle}$. However, spatially varying exponents $h^\pm=h^\pm({\mathbf x})$ can prevent this cancellation when correlated with $s_r$ because of the averaging $\langle . \rangle$. This leads us to multifractal theories~\citep{Carbone1993,PolitanoPouquet1995}, in which such local Hölder scaling\footnote{In hydrodynamic turbulence, Kolmogorov's refined similarity theory relates velocity increments to the locally averaged energy dissipation and its spatial fluctuations \citep{Obukhov1962,Kolmogorov1962}. Multifractal theory describes these fluctuations through local scaling laws $|\delta_r\boldsymbol u|\sim r^{h({\mathbf x})}$, with different H\"older exponents $h({\mathbf x})$ occurring on sets of different fractal dimensions. Averaging yields structure-function exponents that change nonlinearly with statistical order, which is a well-known sign of intermittency \citep{ParisiFrisch1985,Frisch1995}. MHD cascade models extended the description to velocity, magnetic, and Els\"asser increments \citep{Carbone1993}, while intermittency was tied to intense sheetlike structures \citep{PolitanoPouquet1995,MullerBiskamp2000}. Recent simulations support refined-similarity relations for Els\"asser increments \citep{JiangEtAl2026}.} $|\delta_r\boldsymbol z_\perp^\pm|\sim r^{h^\pm({\mathbf x})}$ gives the structure-function exponents $\zeta_p^\pm$, and the value of $\zeta_2^\pm$ determines the energy spectrum; see Eqs.~\eqref{structure} and \eqref{energy}. However, as with Equation~(\ref{eq:weightedhierarchy}), the multifractal description of the Elsässer increments alone does not determine the scale dependence of the weighted angle either. 
  
Our results in the present work are twofold: first, we find analytical and numerical support for some of the established results in the literature discussed above: increments with larger Els\"asser amplitude tend to have smaller individual folded angles and the unweighted mean changes little with scale, whereas the scale dependence becomes stronger as the power of the amplitude weight is increased. The growth of the normalized moments $ I_2(r)={\langle A_r^2\rangle}/{\langle A_r\rangle^2}$ and $I_3(r)={\langle A_r^3\rangle}/{\langle A_r\rangle^3}$ toward smaller separations, together with the comparison of low-order and second-order amplitude statistics, shows that intense increments make an increasing contribution to the measured moments. Since these increments also have smaller angles, intermittency strengthens the scale dependence of the amplitude-weighted angle.

Next, we turn to the question of separating different contributions to the scale-dependence of the weighted angle in an exact way. We use the finite-scale Price equation \citep{Price1970,Price1972,Frank2012Price} to show how the scale dependence of the weighted mean $\langle s_r\rangle_{A}$ can arise  from changes of the unweighted mean $\langle s_r \rangle$ with scale, or from changes in the amplitude weight $A_r$ or correlated changes in both quantities. To study statistical correlations of amplitude and angle, we then employ the joint stochastic process $(\ln A_r, s_r)$ and analyze its conditional moments. 

For numerical and observational tests, we use a high-resolution forced full-MHD simulation from the Johns Hopkins Turbulence Database (JHTDB) \citep{JHTB2,JHTB1,JHTDBMHD}, an independent balanced strong-guide-field reduced-MHD simulation \citep{Beresnyak2014,Beresnyak2015}, and data from NASA's Wind spacecraft \citep{NASA1,NASA2,NASA3,Kovaletal2021,Wind3DPMoments}. In both numerical simulations the decrease of weighted angle results from a negative joint contribution of amplitude and angle changes while the separate angular and amplitude contributions are both non-negative. At fixed initial angles, higher-amplitude increments undergo smaller angular changes, an ordering that can also arise from vector geometry. 


\section{Price decomposition}
\label{sec:Priceequation}

Let us construct Els\"asser increments at $r_>$ and $r_<$  at the same midpoint and with the corresponding labeled direction in the local perpendicular plane at each scale, with $r_> > r_<$ in the inertial range. We write $A_>=A_{r_>}$, $A_<=A_{r_<}$, $s_>=s_{r_>}$and \(s_<=s_{r_<}\). Rearrangement of terms in Equation~\eqref{eq:weightedhierarchy} and a little algebra  gives us
\begin{align*}
 \left\langle s_r^q\right\rangle_{A^p}(r_<)-\left\langle s_r^q\right\rangle_{A^p}(r_>)
 &=\frac{\left\langle A_<^p\left(s_<^q-s_>^q\right)\right\rangle}{\left\langle A_<^p\right\rangle}\\
 +&\frac{\left\langle\left[s_>^q-\left\langle s_r^q\right\rangle_{A^p}(r_>)\right]\left(A_<^p-A_>^p\right)\right\rangle}
 {\left\langle A_<^p\right\rangle}.
\end{align*}
Using $A_<^p=A_>^p+(A_<^p-A_>^p)$, we split the first term and write
\begin{equation}
 \left\langle s_r^q\right\rangle_{A^p}(r_<)-\left\langle s_r^q\right\rangle_{A^p}(r_>)
 =C_\theta^{(p,q)}+C_\times^{(p,q)}+C_A^{(p,q)},
 \label{eq:finiteDecomposition}
\end{equation}
where
\begin{equation}
 \begin{aligned}
 C_\theta^{(p,q)}&:=\frac{\left\langle A_>^p\left(s_<^q-s_>^q\right)\right\rangle}{\left\langle A_<^p\right\rangle},\\
 C_\times^{(p,q)}&:=\frac{\left\langle\left(A_<^p-A_>^p\right)\left(s_<^q-s_>^q\right)\right\rangle}{\left\langle A_<^p\right\rangle},\\
 C_A^{(p,q)}&:=\frac{\left\langle\left[s_>^q-\left\langle s_r^q\right\rangle_{A^p}(r_>)\right]\left(A_<^p-A_>^p\right)\right\rangle}{\left\langle A_<^p\right\rangle}.
 \end{aligned}
 \label{eq:finitePriceterms}
\end{equation}
Equation~\eqref{eq:finiteDecomposition} is the finite-scale Price equation for the $A^p$-weighted angular moment.  For $p=q=1$, with $\Delta A=A_<-A_>$ and $\Delta s=s_<-s_>$, the three terms are denoted $C_\theta$, $C_\times$, and $C_A$.

In Price's original two-term decomposition, with initial weights $A_>$ and relative \emph{fitness} $w=A_</A_>$, $C_\theta+C_\times=\langle A_<\Delta s\rangle/\langle A_<\rangle$ is called the \emph{transmission term} while $C_A$, which measures the amplitude contribution using the larger-scale angles, is called the \emph{selection term}. The term $C_\theta$ measures the angular contribution using the larger-scale amplitudes and $C_\times$ the joint contribution, thus $
C_\theta+C_\times
=
\frac{\langle A_<\Delta s\rangle}
{\langle A_<\rangle}$ represents the amplitude-weighted mean change in \(s\) between the two scales, using the smaller-scale amplitudes as weights.

We note that $A_r$ and $s_r$, which were defined using Els\"asser fields, can be rewritten in terms of the perpendicular velocity and magnetic increments $\delta_r\boldsymbol u_\perp$ and $\delta_r\boldsymbol B_\perp$ and the angle between them, $\alpha_r\in[0,\pi]$. Writing
\begin{align*}
 \mathcal E_r&:=|\delta_r\boldsymbol u_\perp|^2+|\delta_r\boldsymbol B_\perp|^2,\\
 \sigma_r&:=\frac{|\delta_r\boldsymbol u_\perp|^2-|\delta_r\boldsymbol B_\perp|^2}{\mathcal E_r},\\
 \sigma_c&:=\frac{2\,\delta_r\boldsymbol u_\perp\cdot\delta_r\boldsymbol B_\perp}{\mathcal E_r},
\end{align*}
we find
\begin{align*}
 A_r&=\mathcal E_r\sqrt{1-\sigma_c^2}
 =\mathcal E_r\left(\sin^2\alpha_r+\sigma_r^2\cos^2\alpha_r\right)^{1/2},\\
 s_r&=\frac{\sqrt{1-\sigma_r^2}\sin\alpha_r}
 {\left(\sin^2\alpha_r+\sigma_r^2\cos^2\alpha_r\right)^{1/2}},
 \qquad A_r>0.
\end{align*}
Thus with an arbitrarily large angle between magnetic and velocity fluctuations, the Elsasser angle can decrease if the imbalance $|\delta_r\boldsymbol u_\perp|^2-|\delta_r\boldsymbol B_\perp|^2$ increases in magnitude i.e., if $|\sigma_r|$ approaches unity; indeed,  at fixed $\alpha_r\neq0,\pi$, $s_r=1$ at $\sigma_r=0$ and tends to zero as \(|\sigma_r|\rightarrow1\)~\citep{Wicks_2013,Sioulas2024,Schekochihin2022}. Indeed, at fixed $\mathcal E_r$ and $\alpha_r$,
\begin{equation*}
  \frac{\partial A_r^2}{\partial\sigma_r^2}\ge 0,
  \;\quad
  \frac{\partial s_r^2}{\partial\sigma_r^2}\le 0,
\end{equation*}
At fixed $\mathcal E_r$ and $\alpha_r$, fluctuations of $|\sigma_r|$ therefore give a non-positive conditional covariance between $A_r$ and $s_r$. They can thus contribute to the measured negative covariance. The scale dependence of $\langle s_r\rangle_A$ depends on how the joint distribution of $\mathcal E_r$, $\sigma_r$, and $\alpha_r$ changes with $r$; we return to this in Section~\ref{Sec;Spectrum}. 

\begin{table*}[t]
\centering
\caption{\footnotesize Signs of the three finite-scale terms in intervals for which the amplitude-weighted angle decreases.}
\label{table:finitesigns}
\begin{tabular}{lccccc}
\hline\hline
Data
& \shortstack{Total\\intervals}
& \shortstack{Intervals with decreasing\\$\langle s\rangle_A$}
& $C_\theta\geq0$
& $C_A\geq0$
& \shortstack{$C_\times<0$\\and $|C_\times|$ largest}\\
\hline
Full MHD (both subvolumes) & 44 & 41 & 41 & 41 & 41\\
Reduced MHD & 9 & 9 & 9 & 9 & 9\\
\hline
\end{tabular}
\end{table*}

We evaluate Equations (\ref{eq:finiteDecomposition})--(\ref{eq:finitePriceterms}) over the scale intervals used in the angular-change analysis (Section~\ref{sec:angleChange}). Measurements are made at pairs of separations $r_> > r_<$ using the same stored Els\"asser field and midpoint, with the corresponding labeled direction in the local perpendicular plane at each scale.  The full-MHD calculation uses two spatially disjoint $448^3$ subvolumes with 22 intervals each. The reduced-MHD calculation uses the nine scale intervals described in Appendix \ref{app:data}. Table \ref{table:finitesigns} lists the signs of the three terms for the intervals in which $\langle s\rangle_A$ decreases and shows  $C_\theta\geq0$, $C_A\geq0$, and $C_\theta+C_\times<0$. Since
\[
 C_\theta=\frac{\langle A_>\Delta s\rangle}{\langle A_<\rangle},
 \qquad
 C_\theta+C_\times=\frac{\langle A_<\Delta s\rangle}{\langle A_<\rangle},
\]
replacing $A_>$ by $A_<$ reverses the sign of the mean angular change. The difference between these two averages is $C_\times$, while $C_A\geq0$ offsets part of its negative contribution to the change of $\langle s\rangle_A$. Note that if interchanging $r_>$ and $r_<$ leaves the two-scale statistics unchanged, then $\langle s\rangle_{A,<}=\langle s\rangle_{A,>}$ and
\begin{equation*}
 C_\theta=C_A=-\frac{1}{2}C_\times
 =\frac{\langle A_>s_<\rangle-\langle A_>s_>\rangle}{\langle A_>\rangle},
\end{equation*}
so the above sign pattern would follow if $\langle A_>s_<\rangle>\langle A_>s_>\rangle$. At the reference spacing used in Appendix~\ref{appendixpricechecks} the three terms are close to this relation, so the net change is a small remainder of larger opposing terms. 

As a side note, let us also explain why we do not use the continuous version of the Price equation\citep{Price1970,Price1972,Frank2012Price}. If $A_r$ and $s_r$ are  smooth functions of $\chi=\ln(r_0/r)$, then $C_\times/\Delta\chi\rightarrow0$ as $\Delta\chi\rightarrow0$, and Equation~\eqref{eq:finiteDecomposition} reduces to the continuous Price equation,
\[
 \frac{d}{d\chi}\langle s_r\rangle_A
 =\left\langle\frac{d s_r}{d\chi}\right\rangle_A
 +\Cov_A\!\left(s_r,\frac{d\ln A_r}{d\chi}\right),
\]
where $\Cov_A(x,y)=\langle xy\rangle_A-\langle x\rangle_A\langle y\rangle_A$ is the amplitude-weighted covariance. At the finite spacings used here, $C_\times/\Delta\chi$ remains non-negligible (Appendix~\ref{appendixpricechecks}) thus we retain the cross term $C_\times$ and use the finite-scale identity.

\section{Amplitude--Angle distribution}
\label{sec:angleChange}

To study the evolution of amplitude and angle together, we consider the joint distribution of $(a_r,s_r)$ with $a_r=\ln A_r$ (since \(A_r\) may span a broad range) and use three conditional moments\footnote{We do not assume Markovianity or a Fokker--Planck description simply because neither is needed here. But for a more general theoretical description, we note that even a successful Chapman--Kolmogorov test here would establish only coarse-grained Markovianity over the tested separations and state variables. A Fokker--Planck equation requires Markov property and also the Kramers--Moyal expansion to truncate at second order. Testing this condition requires evaluating higher order moments in the limit of vanishing separation which is not accessible in our numerical data.}:
\begin{align*}
 \mu_s(a_>,s_>)&:=\langle s_<\mid a_>,s_>\rangle-\langle s_<\rangle,\\
 M_s^{(2)}(a_>,s_>)&:=\left\langle(s_<-s_>)^2\mid a_>,s_>\right\rangle,\\
 M_\theta^{(2)}(a_>,s_>)&:=\left\langle(\Delta\theta)^2\mid a_>,s_>\right\rangle.
\end{align*}
where $\Delta\theta=\theta_<-\theta_>$. The sign of $\mu_s$ shows whether the smaller-scale values lie below or above the unweighted mean at $r_<$. The two mean-square quantities measure angular change irrespective of direction; if angular change does not depend on initial amplitude at fixed $s_>$, both are constant across the amplitude intervals.

\begin{figure*}
\centering
\includegraphics[width=\textwidth]{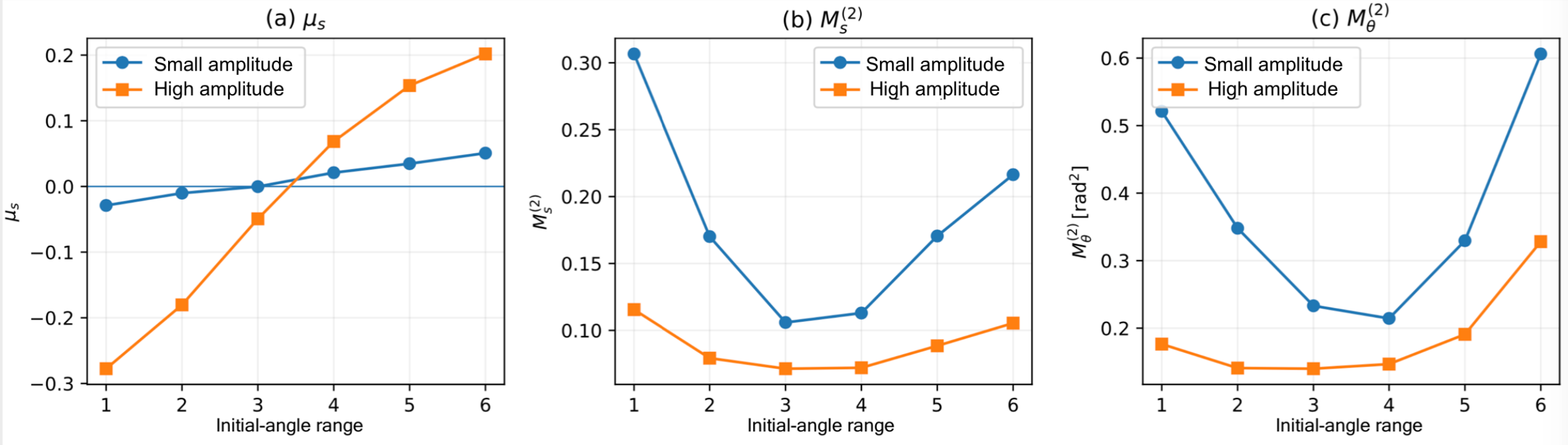}
\caption{\footnotesize Angular change of perpendicular Els\"asser increments in the JHTDB simulation for \(r_>\simeq63.6\rightarrow r_<\simeq55.4\). The measurements are divided into six equal-count intervals of initial angle, ordered from smallest to largest along the horizontal axis, and then into six amplitude intervals. Within each angle interval, the smallest-amplitude one-sixth is shown in blue and the highest-amplitude one-sixth in orange. (a) Mean smaller-scale value relative to the unweighted mean at $r_<$. (b) Mean-square change of $s=\sin\theta$. (c) Mean-square change of the folded angle.}
\label{fig;fullmhdanglechange}
\end{figure*}

At each \(r_>\), we divide the measurements  into $N$ equal-count intervals of $s_>$ and each initial-angle interval is then divided into $N$ equal-count intervals of \(a_>\). Figure \ref{fig;fullmhdanglechange} shows a representative JHTDB transition computed with $N=6$. Within every initial-angle interval, the high-amplitude increments undergo a smaller angular change than the low-amplitude increments (panels b and c). Both measures show the same ordering throughout the full-MHD and reduced-MHD ranges. Appendix~\ref{appendixC:angle_checks} gives the binning and robustness checks.

It is crucial to distinguish a possible dynamical effect in the turbulent cascade from an elementary geometric effect in these results. For a given displacement of its tip, a vector with greater initial magnitude changes direction by a smaller angle. So high-amplitude increments can change angle less, as we go to smaller scales, simply because they are larger vectors. To distinguish this geometric effect from a possible dynamical dependence on amplitude, we construct artificial increments by randomly exchanging their initial amplitudes. Specifically, both Elsässer increments are rescaled by the same factor so that their initial angle and the ratio of their magnitudes remain unchanged. We then apply the original, measured vector changes between \(r_>\) and \(r_<\) to these rescaled increments and recalculate their angles at $r_<$. The decrease of angular change with increasing initial amplitude is even stronger in this calculation than in the actual measurements. Thus, the observed ordering can arise from vector geometry alone; see Appendix~\ref{appendixC:angle_checks}.

\subsection{Time evolution}
\label{sec:time}

Let us also examine the evolution of amplitude and angle in time. At each separation, let HS denote High-amplitude Small-angle increments and HL denote High-amplitude Large-angle increments. We select the highest-amplitude \(10\%\)  (and other values; see Appendix~\ref{appendix4:temporalAnalysis}) ranked by the  $A_r$ and then divide them at the mean folded angle at the same separation. The physical-time comparisons below are made at fixed $r$ and do not enter $C_\theta$, $C_A$, or $C_\times$.

In the JHTDB simulation, we follow increments at the same midpoint and separation in time. For the matched-amplitude comparison, each HL increment is paired with an HS increment having nearly the same initial $A_r$ and we test whether
\begin{align*}
 \left\langle\Delta\ln A_r\right\rangle_{\rm HL}
 &<\left\langle\Delta\ln A_r\right\rangle_{\rm HS},\\
 \Delta\ln A_r
 &=\ln A_r(t+\Delta t)-\ln A_r(t).
\end{align*}
after one, two, and four stored-snapshot intervals. The mean HL-minus-HS change is negative at all three elapsed times and its $95\%$ interval lies below zero after four intervals. The state-loss fractions $D_{\rm HL}$ and $D_{\rm HS}$, defined by failure to retain the initial amplitude and angle conditions, satisfy $D_{\rm HL}>D_{\rm HS}$. Because state loss is threshold dependent, this ratio is reported descriptively and is not used for dynamical inference. Table~\ref{tab_appendix_temporalloss.} and Appendix~\ref{appendix4:temporalAnalysis} give the numerical values and threshold tests. A fixed-$r$ analysis also finds that larger initial amplitudes are associated with smaller subsequent angular changes after controlling for the initial angle and changes of the local geometry (Appendix~\ref{appendix:fixedranglechange}). As in the scale-transition analysis, vector geometry alone can produce this ordering (Appendix~\ref{appendixC:angle_checks}).

The Politano--Pouquet exact relation fixes the sign of its third-order product for transfer toward smaller perpendicular scales. In the reduced-MHD simulation, we evaluate this product separately in intervals of initial amplitude and angle. The highest-amplitude, largest-angle source interval contributes with the forward-transfer sign (Appendix \ref{appendixPolitano}). The same source interval undergoes comparatively small angular changes. In the JHTDB simulation, the local Elsässer advective term $-(\boldsymbol z^\mp\!\cdot\nabla)\boldsymbol z^\pm$ reproduces almost the same dependence of $\Delta\ln A_r$ and $\Delta s_r$ on the initial amplitude and angle (Figure~\ref{figure_app_Nonlinear_Attribution_term}). This agreement includes large-scale sweeping and propagation and does not isolate the scale-local cascade interaction.

\section{Multifractality}
\label{Sec;Spectrum}
%
To separate the condition that fixes the energy spectrum from the condition that fixes the scale dependence of the amplitude-weighted angle, let us first recall the standard multi-fractal picture. Let \(i\) label one midpoint and perpendicular direction measured at both $r_>$ and $r_<$. Over this scale interval, the finite-scale amplitude exponents are
\begin{equation*}
 h_i^\pm=
 \frac{\ln|\delta_{r_<}\boldsymbol z_{\perp,i}^\pm|
       -\ln|\delta_{r_>}\boldsymbol z_{\perp,i}^\pm|}
      {\ln(r_</r_>)}.
\end{equation*}
Equivalently, $|\delta_{r_<}\boldsymbol z_{\perp,i}^\pm|=|\delta_{r_>}\boldsymbol z_{\perp,i}^\pm|(r_</r_>)^{h_i^\pm}$ are finite-scale log ratios. Averaging the $p$th powers gives the structure functions, whose scale dependence defines $\zeta_p^\pm$,
\begin{equation}\label{structure}
 S_p^\pm(r)=\left\langle|\delta_r\boldsymbol z_\perp^\pm|^p\right\rangle
 \propto r^{\zeta_p^\pm}.
\end{equation}
For the two scales;
\begin{equation*}
 \frac{S_p^\pm(r_<)}{S_p^\pm(r_>)}
 =
 \frac{\left\langle
 |\delta_{r_>}\boldsymbol z_\perp^\pm|^p
 \left(\frac{r_<}{r_>}\right)^{p h^\pm}
 \right\rangle}
 {\left\langle|\delta_{r_>}\boldsymbol z_\perp^\pm|^p\right\rangle}.
\end{equation*}
Thus the finite-scale exponents are weighted by the $p$th powers of the corresponding larger-scale amplitudes. For a fixed perpendicular plane, the standard relation gives
\begin{equation}\label{energy}
 E^\pm(k_\perp)\propto k_\perp^{-1-\zeta_2^\pm}.
\end{equation}
(See Appendix~\ref{app.spectra}.)
With $E^\pm$ normalized by $\frac14\langle|\boldsymbol z_\perp^\pm|^2\rangle=\int_0^\infty E^\pm(k_\perp)\,dk_\perp$, the total perpendicular spectrum defined in the Introduction is \(E(k_\perp)=E^+(k_\perp)+E^-(k_\perp)\). When $E^+$ and $E^-$ have the same inertial-range exponent, their sum has the same exponent. In particular, \(\zeta_2^+=\zeta_2^-=2/3\) gives $E(k_\perp)\propto k_\perp^{-5/3}$. Intermittent models use local scaling exponents to describe $\zeta_p^\pm$ \citep{ChandranSchekochihinMallet2015,MalletEtAl2016,MalletSchekochihin2017}.

We now use the finite-scale decomposition of Section~\ref{sec:Priceequation} to show how the same finite-scale exponents enter the weighted-angle change:

\begin{eqnarray}\nonumber
 \frac{A_{<,i}}{A_{>,i}}&=&\left(\frac{r_<}{r_>}\right)^{h_i^++h_i^-}\\\nonumber
 \Delta A_i&=&A_{>,i}\left[\left(\frac{r_<}{r_>}\right)^{h_i^++h_i^-}-1\right].
\end{eqnarray}
Thus the sum of finite-scale amplitude exponents, \(h_i^++h_i^-\), determines the amplitude change between the two scales. Substitution into Equation~(\ref{eq:finitePriceterms}) with $p=q=1$ then yields
\begin{align*}
  C_\theta&=\frac{\langle A_> \Delta s \rangle}{\langle A_<\rangle}\\
  C_\times&=
 \frac{\left\langle
 A_> \Delta s \left[\left(\frac{r_<}{r_>}\right)^{h^++h^-}-1\right]
 \right\rangle}
 {\langle A_<\rangle}\\
  C_A&=
 \frac{\left\langle
 A_>\left(s_>-\langle s\rangle_{A,>}\right)
 \left[\left(\frac{r_<}{r_>}\right)^{h^++h^-}-1\right]
 \right\rangle}
 {\langle A_<\rangle}.
\end{align*}
By Equation~\eqref{eq:finiteDecomposition}, their sum is $\langle s\rangle_{A,<}-\langle s\rangle_{A,>}$. Here $C_\theta$ is the direct angular contribution, while $C_\times$ and $C_A$ contain the correlations of $\Delta A$ with $\Delta s$ and $s_>-\langle s\rangle_{A,>}$, respectively.

We now compare the two conditions with dynamic-alignment phenomenology. According to Equation~(\ref{energy}), $E^\pm(k_\perp)\propto k_\perp^{-3/2}$ requires $\zeta_2^\pm=1/2$, which is a condition on the finite-scale exponents $h_i^\pm$ weighted by the squared larger-scale amplitudes. By contrast, the weighted-angle scaling $\langle s_r\rangle_A\propto r^{1/4}$, which translates into
\[
 \langle s\rangle_{A,<}
 =\left(\frac{r_<}{r_>}\right)^{1/4}\langle s\rangle_{A,>},
\]
in our formalism, requires
\begin{equation}\label{eq.main}
 C_\theta+C_A+C_\times
 =\left[\left(\frac{r_<}{r_>}\right)^{1/4}-1\right]\langle s\rangle_{A,>}.
\end{equation}
This condition concerns the finite-scale amplitude--angle terms, whereas the spectral condition on $h_i^\pm$ is independent of angle, so neither condition implies the other. The $k_\perp^{-3/2}$ energy spectrum can therefore occur with either a scale-independent weighted angle or a scale-dependent $\langle s_r\rangle_A\propto r^{1/4}$. Observing a $k_\perp^{-3/2}$ scaling in simulations is not enough evidence for a cascade-wide dynamic alignment in MHD turbulence.

\begin{figure*}[t]
\centering
\includegraphics[width=0.85\textwidth]{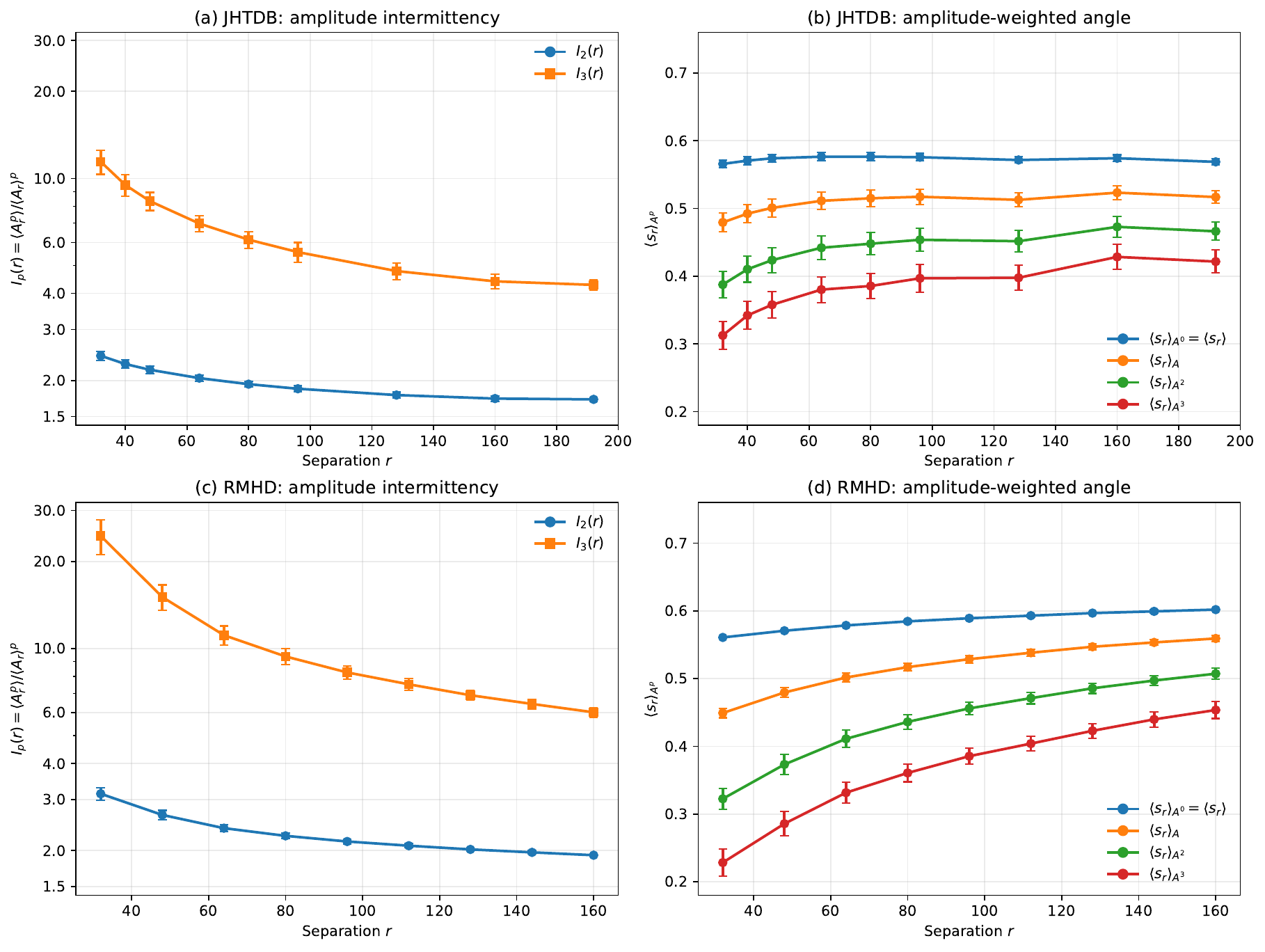}
\caption{\footnotesize Normalized amplitude moments and amplitude-weighted means in full MHD (top) and balanced RMHD (bottom). Panels (a) and (c) show $I_2$ and $I_3$; panels (b) and (d) show $\langle s_r\rangle_{A^p}$ for $p=0,1,2,3$.}
\label{figure;higherMoments}
\end{figure*}

To see why stronger amplitude weighting changes the measured angle, let us now examine how the amplitude distribution changes with scale. The normalized moments
\begin{equation}\label{Eyink1}
 I_p(r):=\frac{\langle A_r^p\rangle}{\langle A_r\rangle^p}
\end{equation}
measure changes in the shape of the amplitude distribution. They are independent of $r$ for a self-similar distribution, while increasing $p$ gives progressively greater weight to the large-amplitude tail~\citep{Caoetal1996,Sreenivasanetal1996,Chenetal2005}. Figure~\ref{figure;higherMoments} shows that $I_2$ and $I_3$ increase toward smaller $r$ in both simulations, while $\langle s_r\rangle_{A^p}$ becomes smaller and more scale dependent as $p$ increases. Thus stronger amplitude weighting selects increments with smaller angles, and the large-amplitude tail contributes increasingly toward smaller scales. At fixed scale, the amplitude--angle covariance is negative in both simulations and in Wind; Figures \ref{fig_app_weighted_covariance} and~\ref{fig:append_wind-data} show the full-MHD and Wind results.

To compare these tail-sensitive moments with the scaling of typical amplitudes, let us use $X_r$ to denote $|\delta_r\boldsymbol z^+_\perp|$, $|\delta_r\boldsymbol z^-_\perp|$, or $A_r^{1/2}$. The slope of $\langle\ln X_r\rangle$ represents the $p\rightarrow0$ behavior, whereas the slope of \(\tfrac12\ln\langle X_r^2\rangle\) is a second-order measure. In the JHTDB simulation, the low-order slope exceeds the second-order slope for all three choices of $X_r$ (Table~\ref{tabel_appen_low_orders}). In the Wind analysis, the same low-order and second-order comparison for $X_\tau=|\delta_\tau\boldsymbol z^\pm|$ gives the same ordering (Appendix \ref{app:windsubsection}). The JHTDB and Wind results therefore show that second-order amplitude statistics are more strongly influenced by intermittent large-amplitude events than the low-order scaling of typical increments. Thus the intermittent large-amplitude tail contributes disproportionately to second-order statistics and is concentrated at smaller angles. Amplitude weighting therefore makes the measured alignment more sensitive to these intense increments than to typical increments.

Why higher amplitude fluctuations are statistically correlated with smaller unweighted angles? As discussed in Section \ref{sec:Priceequation}, smaller unweighted Els\"asser angles can also result if  $|\delta_r\boldsymbol u_\perp|^2-|\delta_r\boldsymbol B_\perp|^2$ increases in magnitude, even when velocity-magnetic fluctuations retain a fixed angle. Using $A_r s_r=\mathcal E_r\sin\alpha_r(1-\sigma_r^2)^{1/2}$ and averaging $\mathcal E_r$ at fixed $(\sigma_r,\alpha_r)$,
\begin{equation}
  \langle s_r\rangle_A=
  \frac{\big\langle \bar{\mathcal E}_r\,\sin\alpha_r\,(1-\sigma_r^2)^{1/2}\big\rangle}
       {\big\langle \bar{\mathcal E}_r\,(\sin^2\alpha_r+\sigma_r^2\cos^2\alpha_r)^{1/2}\big\rangle},
\end{equation}
with $\bar{\mathcal E}_r:=\langle\mathcal E_r\,|\,\sigma_r,\alpha_r\rangle $ so $\mathcal E_r$ acts only as a weight on the $(\sigma_r,\alpha_r)$ statistics. There are three possibilities:

(i) If $\bar{\mathcal E}_r$ is constant and the $(\sigma_r,\alpha_r)$ statistics are scale independent, $\mathcal E_r$ cancels and $\langle s_r\rangle_A$ is scale independent. Figure~\ref{figure;higherMoments} instead shows scale dependence of $\langle s_r\rangle_{A^p}$ that becomes stronger as $p$ increases.\\
(ii) If $\bar{\mathcal E}_r$ is constant but the $(\sigma_r,\alpha_r)$ statistics change with $r$, the amplitude tail plays no role, and the change would generally appear in $\langle s_r\rangle$. But this is not generally supported by simulations and our numerical results which show a nearly constant unweighted angle.\\
(iii) If $\bar{\mathcal E}_r$ increases with $|\sigma_r|$, more energetic increments give greater weight to smaller $s_r$. If this dependence strengthens toward smaller $r$, $\langle s_r\rangle_A$ can decrease while $\langle s_r\rangle$ remains nearly constant. Small angles in the large-amplitude tail may also result from $\alpha_r$ close to $0$ or $\pi$; both effects may act together.

\section{Discussion}
\label{Discussion}

 Previous work has explored relations among the scale dependence of the weighted
Els\"asser angle, the Els\"asser amplitude, and intermittency effects in MHD
turbulence. It has also established how all this is translated into corresponding relations in terms of magnetic and velocity increments. For instance, Els\"asser increments with higher amplitudes are correlated  with smaller unweighted angles $\theta_r$ at any given inertial scale $r$. Unweighted and amplitude-weighted Elsässer angles scale differently with separation  and the scaling also depends on  the amplitude weighting and moment order.

However, a clearer theoretical understanding requires both an exact separation of the different contributions to this
scale dependence and a characterisation of the joint statistics of amplitude and
angle as a function of scale. In this paper, we have taken first steps toward
this goal, using analytical arguments based on the Price equation from evolutionary biology \citep{Price1970,Price1972,Frank2012Price} and the joint
distribution of amplitude and angle $(A_r, s_r)$, together with two publicly
available high-resolution simulations and NASA's Wind data, to examine the scale
dependence of $\langle s_r\rangle_A = \langle A_r s_r\rangle/\langle A_r\rangle$,
with $A_r = |\delta_r\boldsymbol z_\perp^+|\,|\delta_r\boldsymbol z_\perp^-|$ and $s_r = \sin\theta_r$.

 The finite-scale Price equation  separates the change of $\langle s_r\rangle_A$ into an angular contribution $C_\theta$, an amplitude contribution $C_A$, and an amplitude-angle joint contribution $C_\times$. In every interval where the weighted angle decreases in both simulations, the joint term $C_\times$ is negative, while the angular term $C_\theta$ and amplitude term $C_A$ are both positive. These terms are individually much larger than their sum, therefore the measured decrease is the small difference left by their cancellation. The scale dependence of the weighted angle is therefore controlled by the joint evolution of amplitude and angle across the same increments. We observe that at any fixed initial angle, larger Els\"asser increments undergo smaller angular changes in both simulations but this should not be misunderstood as a purely dynamic effect; our numerical tests show that vector geometry alone can also reproduce this ordering. This should be expected, in fact, because for the same displacement of a vector tip, the larger vector turns through a smaller angle. The corresponding ordering in the time-dependent measurements can also arise from the same geometrical effect.

The velocity and magnetic variables give a more detailed interpretation of these changes. Let $\alpha_r$ denote the angle between the perpendicular velocity and magnetic increments. The normalized residual energy $\sigma_r$ measures the difference between their squared magnitudes relative to their sum, while $
\mathcal E_r=
|\delta_r\boldsymbol u_\perp|^2+
|\delta_r\boldsymbol B_\perp|^2$
measures their combined squared increment level. The standard geometrical relations show that both $A_r$ and $s_r$ depend on $\alpha_r$ and on $|\sigma_r|$, while $\mathcal E_r$ enters only $A_r$ \citep{Wicks_2013,SioulasEtAl2024}. Consequently, $\Delta s$ combines reorientation of the velocity and magnetic increments with a change in their magnitude ratio. The change $\Delta A$ contains both of these effects together with the change in $\mathcal E_r$. The Price terms therefore describe how velocity--magnetic reorientation, changes in the relative magnitudes, and changes in the overall increment level combine between two separations. Resolving $\Delta s$ into its $\alpha_r$ and $|\sigma_r|$ parts would separate the first two effects.

It remains to ask why higher-amplitude Els\"asser increments have smaller unweighted angles. With $I_p$ defined in Equation~(\ref{Eyink1}), Figure~\ref{figure;higherMoments} shows that $I_2$ and $I_3$ increase toward smaller $r$, while $\langle s_r\rangle_{A^p}$ becomes smaller and more scale dependent as $p$ increases. Thus stronger amplitude weighting selects increments with smaller angles. For each increment,
\[
 \bigl|\delta_r\boldsymbol z_\perp^+\!\cdot\delta_r\boldsymbol z_\perp^-\bigr|
 =\bigl||\delta_r\boldsymbol u_\perp|^2-|\delta_r\boldsymbol B_\perp|^2\bigr|
 =A_r\cos\theta_r .
\]
The correlation between large $A_r$ and small $\theta_r$ therefore means that
\[
 \frac{\bigl||\delta_r\boldsymbol u_\perp|^2-|\delta_r\boldsymbol B_\perp|^2\bigr|}{A_r}
\]
increases on average with $A_r$. The relations in Section~\ref{sec:Priceequation} show that a small $\theta_r$ can result from large $|\sigma_r|$, from $\alpha_r$ close to $0$ or $\pi$ when the velocity and magnetic increments have unequal magnitudes, or from both.

At the largest inertial-range separations considered here, higher-amplitude Els\"asser increments already have smaller unweighted angles, in agreement with earlier fixed-scale measurements \citep{BeresnyakLazarian2006,MalletSchekochihinChandran2015,MalletEtAl2016}. The same correlation remains present toward the smallest separations analysed. Figure~\ref{fig;fullmhdanglechange} and Appendix~\ref{appendixC:angle_checks} show our additional result that, at fixed initial angle, higher-amplitude increments undergo smaller angular changes toward smaller separations. These results, together with the transfer of energy from larger to smaller scales, suggest that the correlation between higher amplitudes and smaller angles may be established by the forcing, or by the fluctuations present at the outer scale, and maintained through the inertial range. This is consistent with the finding of \citet{Beresnyak2012} that the maximum alignment slope is tied to the outer scale.

The full-MHD run suggests how the forcing can establish this correlation. The JHTDB simulation uses a steady Taylor--Green force at $k_f=2$, a fixed large-scale force applied only to $\boldsymbol u$ \citep{JHTDBMHD}. The same forcing term therefore enters the two Els\"asser equations with the same sign. Near the forcing scale, if a large common velocity contribution dominates the increment, then
\[
 \delta_r\boldsymbol z_\perp^+
 \simeq
 \delta_r\boldsymbol z_\perp^-
 \simeq
 \delta_r\boldsymbol u_\perp .
\]
The two Els\"asser increments are then nearly parallel, $A_r\simeq|\delta_r\boldsymbol u_\perp|^2$ is large, and $\theta_r$ is small.

In the RMHD simulation, the forces on $\boldsymbol z^+$ and $\boldsymbol z^-$ are independent and restricted to the lowest wavenumbers $k<1.42$ \citep{Beresnyak2012,Beresnyak2014}. The Elsässer fluctuations lie in the plane perpendicular to the guide field, whose direction is denoted by $\hat{\boldsymbol b}_0$. For a given perpendicular wavevector $\boldsymbol k_\perp$, incompressibility requires
\[
\boldsymbol k_\perp\cdot\boldsymbol z^\pm_{\perp,\boldsymbol k}=0.
\]
Since both Fourier components lie in the same two-dimensional plane, this constraint leaves only one allowed direction, \(\hat{\boldsymbol b}_0\times\hat{\boldsymbol k}_\perp\). Thus, the two components at that wavevector have the form $a_{\boldsymbol k}(\hat{\boldsymbol b}_0\times\hat{\boldsymbol k}_\perp)$ and $b_{\boldsymbol k}(\hat{\boldsymbol b}_0\times\hat{\boldsymbol k}_\perp)$, respectively, where the coefficients can differ in magnitude and sign. They therefore lie along the same line and are either parallel or antiparallel. Since taking a spatial increment does not change the polarization direction of an individual Fourier mode, near the forcing scale, if a high-$A_r$ increment is dominated by the same low-wavenumber mode in both fields, the two increments are parallel or antiparallel and the folded angle is small.

The two forcing schemes can therefore produce high-amplitude, small-angle increments near the forcing scale in different ways: through the common velocity contribution in full MHD and through the common direction of each forced mode in reduced MHD. The smaller angular changes of higher-amplitude increments can help maintain this correlation toward smaller scales. Figure~\ref{figure;higherMoments} shows that the large-amplitude tail contributes increasingly as $r$ decreases, while the negative $C_\times$ lowers $\langle s_r\rangle_A$ in the measured intervals. The contribution of high-amplitude, small-angle increments to the weighted angle therefore becomes scale dependent. The Wind measurements show the same correlation in observed turbulence. Simulations that change the forcing while keeping the remaining setup fixed would test the conjecture.

Whatever establishes the correlation between high amplitude and small angle, it does not determine the energy spectrum. The multifractal description gives the corresponding amplitude picture. The finite-scale exponents $h^\pm$ determine how the two Els\"asser amplitudes change between separations and, through their second-order statistics, determine the perpendicular spectrum. The weighted-angle scaling also depends on the angle and on its correlations with those amplitude changes. A $k_\perp^{-3/2}$ spectrum is therefore compatible with either a nearly constant $\langle s_r\rangle_A$ or a scaling such as $\langle s_r\rangle_A\propto r^{1/4}$. Intermittent RMHD theories connect intense fluctuations, smaller angles, and greater anisotropy within the perpendicular plane \citep{ChandranSchekochihinMallet2015,MalletEtAl2016,MalletSchekochihin2017,Schekochihin2022} while our results further show how the increasing statistical weight of those intense fluctuations enters the measured alignment scaling. An alignment measurement aimed at nonlinear depletion should include the unweighted angle together with $C_\theta$, $C_\times$, and $C_A$. These quantities distinguish a population-wide change of angle from the increasing influence of intermittent, high-amplitude increments.

\section*{Acknowledgments}
I thank E. Vishniac for discussions of dynamic-alignment measurements and the roles of Els\"asser-amplitude weighting and conditioning on current density. I thank A. Beresnyak for comments that motivated threshold, Fourier-randomization, and reduced-MHD checks, A. Lazarian for discussions of polarization intermittency, and G. L. Eyink for suggesting the low-order amplitude check and for comments on the weighting of high-amplitude increments. I also thank the anonymous referee, whose comments helped sharpen the paper and make it more concise. ChatGPT Pro was used to help debug Python code and edit the text for readability.

\section*{Data Availability}
The full-MHD simulation data are publicly available through the Johns Hopkins Turbulence Database (JHTDB). The reduced-MHD fields are from the public MHD datacubes described by \citet{Beresnyak2014,Beresnyak2015}. Wind data are public NASA CDAWeb products \citep{NASA1,NASA2,NASA3,Kovaletal2021,Wind3DPMoments}. The analysis scripts and  metadata are archived at Zenodo under https://doi.org/10.5281/zenodo.23813388.

\appendix
\renewcommand{\theHequation}{appendix.\Alph{section}.\arabic{equation}}

\section{Perpendicular structure functions and energy spectra}
\label{app.spectra}

The relations below are standard, but we state them because the connection
between the second-order structure function and the perpendicular energy
spectrum is central to our discussions. Throughout this appendix, $k_\perp$ is
defined relative to a fixed direction, such as an imposed guide field or a
global mean field. If no fixed direction exists, there is no unique global
Fourier $k_\perp$; the direction-independent one-dimensional global Fourier
spectrum is then the shell-integrated spectrum $E(k)$. A perpendicular statistic
relative to a scale-dependent local magnetic field must instead be defined by a
spatially localized construction, such as a structure function or wavelet
transform. The second-order structure function is
\begin{equation*}
  S_2^\pm(r)=\left\langle|\delta_r\boldsymbol z_\perp^\pm|^2\right\rangle
 \propto r^{\zeta_2^\pm}.
\end{equation*}
The total perpendicular spectrum used in the Introduction is $E(k_\perp)=E^+(k_\perp)+E^-(k_\perp)$. Here $E^\pm(k_\perp)$ is one quarter of the trace of the perpendicular Els\"asser spectral tensor, integrated over $k_\parallel$ and over the azimuth of $\boldsymbol k_\perp$ at fixed $k_\perp$. The spectra are normalized by
$$
 \frac14\left\langle|\boldsymbol z_\perp^\pm|^2\right\rangle
 =\int_0^\infty E^\pm(k_\perp)\,d k_\perp.
$$

For the angular average over separation directions in the perpendicular
plane, statistical homogeneity gives
$$
 S_2^\pm(r)
 =8\int_0^\infty E^\pm(k_\perp)
 \left[1-J_0(k_\perp r)\right]d k_\perp,
$$
where $J_0$ is the Bessel function of the first kind of order zero. The factor $8$ follows from
the normalization above, and $J_0$ from the angular average of the Fourier
phase. If the direction within the perpendicular plane is retained, the
corresponding angular dependence must also be retained.

The Fourier relation above therefore applies only to a fixed perpendicular
plane, as in the RMHD run. The JHTDB simulation has no imposed guide field, and
our analysis there uses structure functions with separation directions
perpendicular to the scale-dependent local field
(Equation~\ref{equation_app_increments}); it does not define a local-field
Fourier spectrum. The measured structure function mixes second- and
higher-order statistics, so the relation above is not an exact Fourier identity
\citep{Matthaeus2012}.

To find the spectral exponent corresponding to $S_2^\pm(r)\propto r^{\zeta_2^\pm}$, let $x=k_\perp r$, so that
$k_\perp=x/r$ and \(d k_\perp=d x/r\). The spectral relation becomes
\begin{equation}\nonumber
 S_2^\pm(r)
 =\frac{8}{r}\int_0^\infty E^\pm(x/r)
 \left[1-J_0(x)\right]d x.
\end{equation}
For a power-law spectrum, the dependence on $r$ is therefore contained in
$E^\pm(x/r)/r$. Matching this dependence to
$S_2^\pm(r)\propto r^{\zeta_2^\pm}$ gives
$$
 E^\pm(k_\perp)\propto k_\perp^{-1-\zeta_2^\pm}.
$$
Substitution gives
\[
 S_2^\pm(r)
 \propto
 r^{\zeta_2^\pm}
 \int_0^\infty x^{-1-\zeta_2^\pm}
 \left[1-J_0(x)\right]d x.
\]
For \(0<\zeta_2^\pm<2\), the remaining integral is finite and independent of
$r$, so the structure function is spectrally local around $k_\perp\sim1/r$.

\section{Numerical Tests}\label{app:data}

Unless stated otherwise, all simulation separations $r$ are given in grid points. Also, angles use in both MHD and RMHD simulations are planar, whereas the Wind analysis uses full three-dimensional increments. Their random-angle baselines therefore differ and we compare scale trends rather than absolute angle levels across datasets.

The full-MHD measurements use the public forced incompressible
\texttt{mhd1024} data set in the Johns Hopkins Turbulence Database
\citep{JHTDBMHD,JHTB1,JHTB2,Eyinketal2013}. The simulation is periodic,
with a $1024^3$ pseudospectral grid, Taylor--Green forcing at $k_f=2$,
$\nu=\eta=1.1\times10^{-4}$, and no imposed uniform guide field. Its
Taylor-scale Reynolds number is approximately $186$.

At each separation, the magnetic field is smoothed with a three-dimensional
Gaussian filter of standard deviation $r/2$ and evaluated at the midpoint.
The direction of this local field is denoted by
$\widehat{\boldsymbol b}_L$, and the projector onto its perpendicular plane is
\begin{equation}
 \mathsf P_L=\mathsf I-
 \widehat{\boldsymbol b}_L\widehat{\boldsymbol b}_L^{\mathsf T}.
 \label{eq:app_projector}
\end{equation}
The centered perpendicular increments are
\begin{equation}
 \delta_r\boldsymbol z^\pm_\perp
 =\mathsf P_L\!\left[
 \boldsymbol z^\pm(\boldsymbol x+\boldsymbol r/2)
 -\boldsymbol z^\pm(\boldsymbol x-\boldsymbol r/2)
 \right],
 \qquad \boldsymbol r\cdot\widehat{\boldsymbol b}_L=0.
 \label{equation_app_increments}
\end{equation}
Off-grid endpoint values are obtained by trilinear interpolation.

For comparisons between separations, the same stored field, midpoint, and
direction index are used. A fixed reference vector at each midpoint assigns
the direction index consistently as the local perpendicular plane changes
with $r$, while the local magnetic field, perpendicular plane, separation
direction, and increment endpoints are recalculated. Midpoints are sampled
with a boundary margin that keeps both centered endpoints inside the
subvolume at every separation. Reference vectors are accepted only if
their projection onto every local perpendicular plane has norm at least
$0.20$; measurements with a non-finite or vanishing local field or projected
Els\"asser increment are discarded.

The fixed-scale analysis uses fifteen randomly selected, non-overlapping
\(320^3\) subvolumes from different stored times, with $120,000$ midpoints in
each. At each midpoint, increments are constructed along eight directions at
\(45^\circ\) intervals in the local perpendicular plane. Because the increments
are centered, opposite directions give identical $A_r$ and $s_r$, so these
are four independent axes, each counted twice. The finite-spacing calculation
in Appendix \ref{appendixpricechecks} uses two $320^3$ subvolumes with $2000$
sampled midpoints in each.

The scale-transition analysis uses two spatially disjoint \(448^3\)
subvolumes. Cube A spans the one-based grid coordinates
$x,y,z=289,\ldots,736$; Cube B is its periodic half-box translation, with
wrapped range $801,\ldots,1024,1,\ldots,224$ in each coordinate. Both cubes
are sampled in the same nine stored fields, with \(4000\) midpoints and eight
distinct unoriented axes at $22.5^\circ$ intervals in the local perpendicular
plane. The 25 separations are logarithmically spaced from $r=96$ to \(r=32\).
Each transition spans three steps of this logarithmic separation sequence, so
$r_</r_>= 3^{-1/8}$. The
finite-scale decomposition (Section~\ref{sec:Priceequation}) and the vector-geometry
control use both cubes; the angular-change analysis
(Section \ref{sec:angleChange}) uses Cube A.

The reduced-MHD test uses the balanced strong-guide-field $1024^3$
ordinary-viscosity run \texttt{b1024n} described by
\citet{Beresnyak2014,Beresnyak2015}. The velocity and magnetic fields are
transverse to the imposed guide field, so the perpendicular plane is the
fixed $y$--$z$ plane. The primary stored field is R108000, with R162000 as a
second-field check. The scale intervals are
$$
 r_>=32,48,64,80,96,112,128,144,160,
 \qquad r_<=\frac{7}{8}r_>.
$$
Two interlaced midpoint lattices provide $524,288$ measurements for each
perpendicular direction. Five stored fields are used for the fixed-scale and
Politano-Pouquet measurements. The RMHD calculation uses two fixed in-plane
directions and therefore assumes in-plane isotropy rather than performing an
explicit continuous angular average.

\subsection{Finite-spacing and matching checks}
\label{appendixpricechecks}

We evaluate $C_\theta$, $C_A$, and $C_\times$ for three scale intervals $h=\Delta\chi$.
Table \ref{tabel_finite_price} gives the terms per unit $h$. For $h_0/4$ and $h_0/16$, the separation changes by about
$0.07$--$1.8$ grid spacings over $32\leq r\leq192$ so these rows depend on
trilinear interpolation between grid points.

\begin{table*}[!t]
\centering
\caption{\footnotesize Finite-spacing decomposition in two $320^3$
full-MHD subvolumes. Each value is averaged over eleven logarithmically
spaced separations in the inertial range $32\leq r\leq192$ within each subvolume and then over
the two subvolumes; the uncertainty is the standard error across the two
subvolume values. The reference interval $h_0=\ln(6)/48$ is the spacing of
49 logarithmically spaced separations over this range.}
\label{tabel_finite_price}
\begin{tabular}{ccccc}
\hline\hline
Scale interval & $\Delta\langle s\rangle_A/h$ & $C_\theta/h$ & $C_A/h$ & $C_\times/h$\\
\hline
$h_0$    & $-0.0045\pm0.0036$ & $+0.1710\pm0.0558$
         & $+0.1554\pm0.0168$ & $-0.3310\pm0.0690$\\
$h_0/4$  & $+0.0678\pm0.0294$ & $+0.2018\pm0.1000$
         & $+0.1089\pm0.0178$ & $-0.2429\pm0.0884$\\
$h_0/16$ & $+0.0433\pm0.0205$ & $+0.0938\pm0.0965$
         & $+0.0462\pm0.0271$ & $-0.0968\pm0.0490$\\
\hline
\end{tabular}
\end{table*}

At every tested spacing, the central value of
$|C_\times/h|$ exceeds $|\Delta\langle s\rangle_A/h|$, so the joint term
remains important relative to the net change. The positive totals at
$h_0/4$ and $h_0/16$ occur when the separation change is below two grid
spacings, where the calculation depends more strongly on trilinear
interpolation. Filtering the local magnetic field in a finite subvolume can
also affect the separate terms through the treatment of the subvolume
boundary.

\subsection{Amplitude dependence of angular change}
\label{appendixC:angle_checks}

Both of second order moments $M_s^{(2)}$ and $M_\theta^{(2)}$ are
smaller in the highest-amplitude interval than in the lowest-amplitude
interval. Their median highest-to-lowest ratios are 0.49 and 0.50,
respectively. The ordering persists for \(4\leq N\leq12\), after excluding
increments below the $5th, 10th,$ or $20th$ percentiles, and when the
smaller-scale increment is constructed using the larger-scale perpendicular
plane and separation direction.

For the representative transition, we let $\boldsymbol q^\pm_>$ and $\boldsymbol q^\pm_<$ be the two projected Elsässer increments at the larger and smaller separations, and $\Delta\boldsymbol q^\pm=\boldsymbol q^\pm_<-\boldsymbol q^\pm_>$ be their measured changes. Within each cube--stored-field sample, initial-angle interval, direction index, and interval of the amplitude ratio $\ln(|\boldsymbol q^+_>|/|\boldsymbol q^-_>|)$, we randomly assign to each measurement the overall initial amplitude of another measurement. If measurement $i$ receives $A_{>,\pi(i)}$, both initial increments are multiplied by
\begin{equation*}
 c_i=\left(\frac{A_{>,\pi(i)}}{A_{>,i}}\right)^{1/2},
 \quad
 \widetilde{\boldsymbol q}^{\,\pm}_{>,i}=c_i\boldsymbol q^\pm_{>,i},
 \quad
 \widetilde{\boldsymbol q}^{\,\pm}_{<,i}
 =\widetilde{\boldsymbol q}^{\,\pm}_{>,i}+\Delta\boldsymbol q^\pm_i.
\end{equation*}
This changes the overall amplitude of the two initial increments but keeps their directions, mutual angle, \(z^+/z^-\) amplitude ratio, and measured changes \(\Delta\boldsymbol q^\pm\) unchanged. Across the cube--stored-field samples, the median ratio
$M_\theta^{(2)}(\text{highest amplitude})/
M_\theta^{(2)}(\text{lowest amplitude})$ is $R_{\rm obs}=0.41$ in the
measured data and $R_{\rm test}=0.32$ after amplitude reassignment. After
excluding the lowest-amplitude \(5\%\) of individual increments, the
corresponding medians are $0.46$ and $0.35$.

Figure \ref{fig_RMHDanglechange} shows one RMHD scale interval. In the
reduced-MHD data, both angular-change measures decrease with amplitude in the
two perpendicular directions, in spatial octants, and in a second stored
field. The perpendicular plane is recalculated at every separation in full
MHD and fixed by the guide field in RMHD.

\begin{figure*}[t]
\centering
\includegraphics[width=\textwidth]{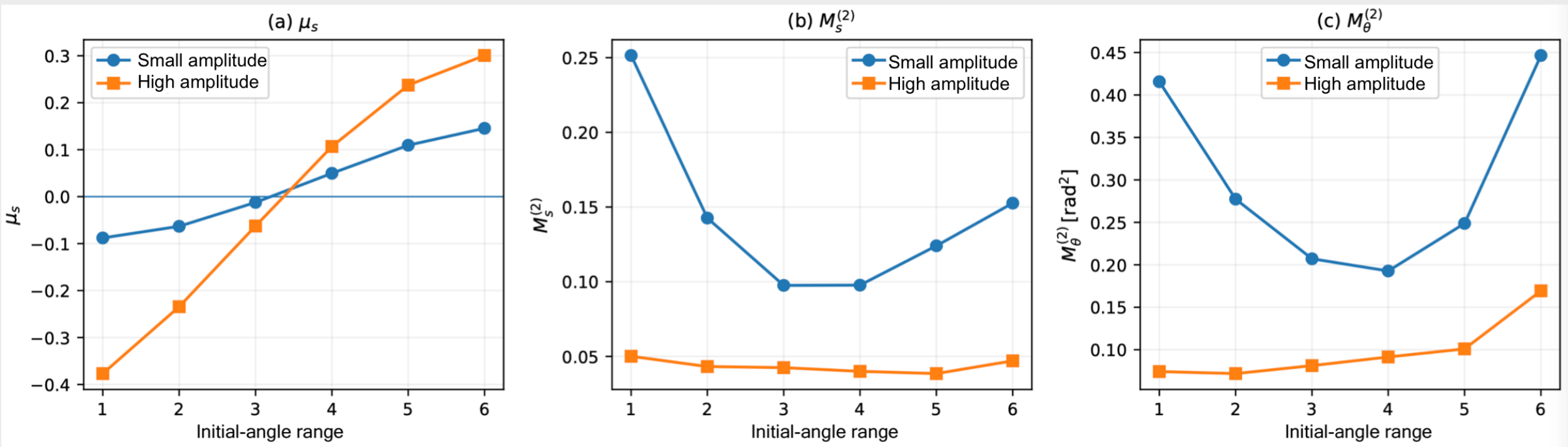}
\caption{\footnotesize Change of angles in the RMHD simulation for $r_>=64\rightarrow r_<=56$. The panels show the same quantities as Figure~\ref{fig;fullmhdanglechange}.}
\label{fig_RMHDanglechange}
\end{figure*}

\subsection{Finite-time amplitude change}
\label{appendix4:temporalAnalysis}

The calculation uses twenty independent $320^3$ full-MHD windows at
\(r=32,48,64,80,96,128,160\), and $192$, each followed through five
consecutive stored snapshots. Adjacent snapshots are separated by
\(\delta t_{\rm snap}=2.5\times10^{-3}\) simulation units. The complete stored interval, $0\leq t\leq2.56$, is about one large-eddy
turnover time \citep{JHTDBMHD}, so the one, two, and four snapshot lags correspond
approximately to $10^{-3}$, $2\times10^{-3}$, and \(4\times10^{-3}\) turnover
times. For either HS and HL states, defined in Subsection \ref{sec:time}, we let $P_{i\rightarrow i}$ denote the fraction of increments that
still satisfy its initial amplitude and angle conditions after a given lag,
and define $D_i=1-P_{i\rightarrow i}$. The ordering
\(D_{\rm HL}>D_{\rm HS}\) holds across all windows and separations and
is unchanged for high-amplitude thresholds of $5\%$, $15\%$, and $20\%$.

The
one-to-one matching described in Subsection \ref{sec:time} pairs almost
all HL measurements (\(99.9\%\)) with HS measurements whose initial
$\ln A_r$ differs by less than $10^{-3}$ on average.
Table~\ref{tab_appendix_temporalloss.} reports the state-loss ratios and the
matched-amplitude changes.

\begin{table*}[t]
\centering
\caption{\footnotesize HL-HS state loss ratios and matched-amplitude changes for HL and HS increments. The third column uses one-to-one matching at nearly equal initial amplitude. Uncertainties are standard errors across twenty windows; the final column gives percentile-bootstrap $95\%$ intervals.}
\label{tab_appendix_temporalloss.}
\begin{tabular}{cccc}
\hline\hline
Elapsed snapshot intervals & $\langle D_{\rm HL}/D_{\rm HS}\rangle$
& $\langle\Delta\ln A_r\rangle_{\rm HL}-\langle\Delta\ln A_r\rangle_{\rm HS}$ & 95\% interval\\
\hline
1 & 2.10 & $-6.54\times10^{-5}\pm5.70\times10^{-5}$
  & $[-1.80\times10^{-4},\,3.90\times10^{-5}]$\\
2 & 2.12 & $-1.88\times10^{-4}\pm1.31\times10^{-4}$
  & $[-4.57\times10^{-4},\,4.22\times10^{-5}]$\\
4 & 2.12 & $-7.44\times10^{-4}\pm2.76\times10^{-4}$
  & $[-1.29\times10^{-3},\,-2.39\times10^{-4}]$\\
\hline
\end{tabular}
\end{table*}

\subsection{Angular change at fixed separation}
\label{appendix:fixedranglechange}

A separate analysis uses 65 consecutive Els\"sser fields separated by
$\Delta t=0.04$ in simulation units. At every stored time, the local magnetic
field, perpendicular plane, separation direction, and increment endpoints
are recalculated. Partial Spearman rank correlations measure the relation
between the initial amplitude $\ln A_r$ and the later angular changes
$|s_1-s_0|$ and $|\theta_1-\theta_0|$, controlling for the initial angle and
changes in the local geometry. Table \ref{tab:timeanglechange} gives
the median correlations over the 64 time intervals. Both correlations are
negative throughout the four separations and remain negative within each of
six initial-angle intervals, after excluding increments below the 5th, 10th,
or 20th percentile, and in a spatially disjoint subvolume.

\begin{table}[t]
\centering
\caption{\footnotesize Median partial Spearman rank correlations between the initial amplitude $\ln A_r$ and the later changes $|s_1-s_0|$ and $|\theta_1-\theta_0|$. The correlations control for the initial angle and changes in the local perpendicular geometry. Negative values correspond to smaller angular changes at higher amplitude.}
\label{tab:timeanglechange}
\begin{tabular}{ccc}
\hline\hline
$r$ & $\rho_s$ & $\rho_\theta$\\
\hline
64  & $-0.346$ & $-0.357$\\
80  & $-0.340$ & $-0.352$\\
96  & $-0.327$ & $-0.343$\\
112 & $-0.318$ & $-0.334$\\
\hline
\end{tabular}
\end{table}

\subsection{Local Els\"asser advective term}

For the advective-term attribution, the later field is evaluated at the
initial endpoints and in the initial perpendicular plane. The local
Els\"asser advective term and its increment between the two endpoints are
\begin{equation}
 \boldsymbol N^\pm=-(\boldsymbol z^\mp\!\cdot\nabla)\boldsymbol z^\pm,
 \qquad
 \boldsymbol R^\pm_{\rm NL}
 =\boldsymbol N^\pm(\boldsymbol x_+)-\boldsymbol N^\pm(\boldsymbol x_-).
 \label{eq:appendix-nonlinearity}
\end{equation}
A trapezoidal update combines the terms measured from the actual fields at
the beginning and end of each stored-time interval. The update estimates the
advective contribution within the complete MHD evolution. The destination-time
term is evaluated from the fully evolved field; pressure, dissipation,
forcing, and numerical error contribute to the residual.

At fixed Eulerian points, the advective term contains sweeping by the
large-scale velocity and propagation along the large-scale magnetic field, as
well as scale-local distortion. The comparison does not isolate the
scale-local cascade interaction. For the primary $10\times10$ partition of initial amplitude and angle, we
compare the measured and advective-update transition kernels using the
source-weighted total-variation distance. Figure~\ref{figure_app_Nonlinear_Attribution_term}
shows this comparison and the correlations between the measured and estimated
conditional mean changes.

\begin{figure*}[t]
\centering
\includegraphics[width=0.82\textwidth]{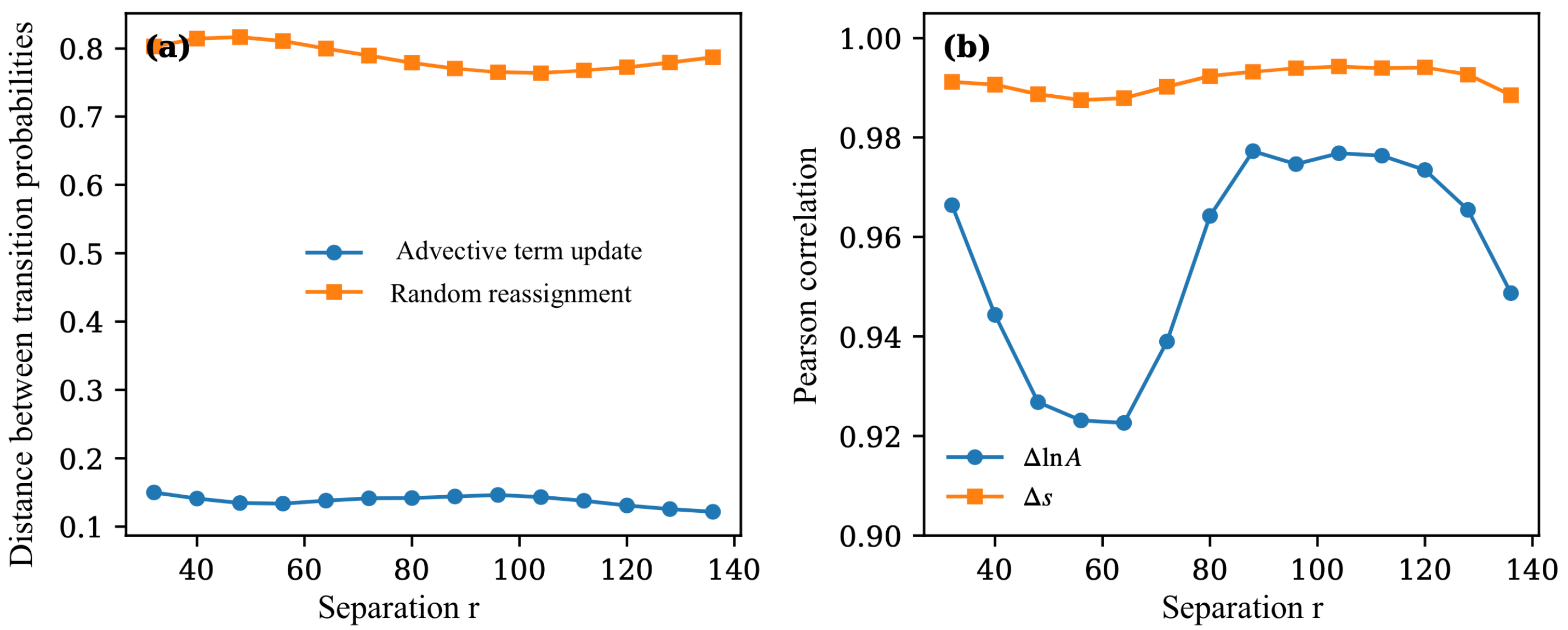}
\caption{\footnotesize Comparison of the measured time change with the change estimated from the local Els\"asser advective term in the full-MHD simulation. (a) Source-weighted total-variation distance between the measured and estimated transitions, together with the random-reassignment control. (b) Pearson correlation between the measured and estimated mean changes of $\ln A$ and $s$ in the initial amplitude--angle intervals.}
\label{figure_app_Nonlinear_Attribution_term}
\end{figure*}

\subsection{Politano-Pouquet check}
\label{appendixPolitano}

For homogeneous, stationary, incompressible MHD in the inertial range, the Politano--Pouquet exact relation is
\citep{PolitanoPouquet1998}
\begin{equation}
 \nabla_{\boldsymbol r}\!\cdot
 \left\langle
 \delta\boldsymbol z^\mp
 \left|\delta\boldsymbol z^\pm\right|^2
 \right\rangle
 =-4\varepsilon^\pm.
 \label{eq:app_pp_general}
\end{equation}
For perpendicular-plane isotropy in RMHD, its integrated form is
\citep{Galtier2011}
\begin{equation}
 \left\langle
 \delta z^\mp_{L,\perp}
 \left|\delta\boldsymbol z^\pm_\perp\right|^2
 \right\rangle
 =-2\varepsilon^\pm r_\perp.
 \label{eq:app_pp_rmhd}
\end{equation}
where
$\delta z^\mp_{L,\perp}=\delta\boldsymbol z^\mp_\perp\!\cdot
\widehat{\boldsymbol r}_\perp$
is the component along the separation direction.

Our RMHD measurements are divided into initial-angle and amplitude intervals
as in Section \ref{sec:angleChange}. For each interval, we form the mean
signed third-order product and the mean-square angular change to the smaller
separation. The exact relation applies to the full ensemble; the conditional
products are used here only as a sign check. The full-sample moment has the forward-transfer sign in $1327$ of
\(1360\) estimates (two perpendicular directions, all sampled separations, and
eight spatial octants). In the interval containing the largest initial angles and highest amplitudes,
the mean of the $+$ and $-$ products is negative at every larger separation,
in every stored field, and for the $4\times4$, $6\times6$, and $8\times8$
divisions of the amplitude-angle plane. Within each initial-angle interval,
the highest-amplitude interval has smaller $M_\theta^{(2)}$ in 3218 of 3240
comparisons.

\subsection{Single scale results}
\label{lastappendix:supporting_Wind}

Figure \ref{fig_app_weighted_covariance} shows the fixed-scale relation between amplitude and angle. The amplitude-weighted mean lies below the unweighted mean on every inertial scale and the normalized covariance is negative in all fifteen subvolumes at all nine separations. Shuffling $A_r$ among the measured angles removes the covariance and makes the amplitude-weighted mean nearly coincide with the unweighted mean. Selecting the highest-amplitude \(10\%\) by $A_r$ lowers the folded mean angle by about $9^\circ$--$13^\circ$ relative to the full sample. Selecting the largest 10\% by current-density magnitude leaves the angle very close to the full-sample curve; Figure~\ref{fig_app_current_densityweighting_Vishniac} shows this comparison.

\begin{figure*}[t]
\centering
\includegraphics[width=0.86\textwidth]{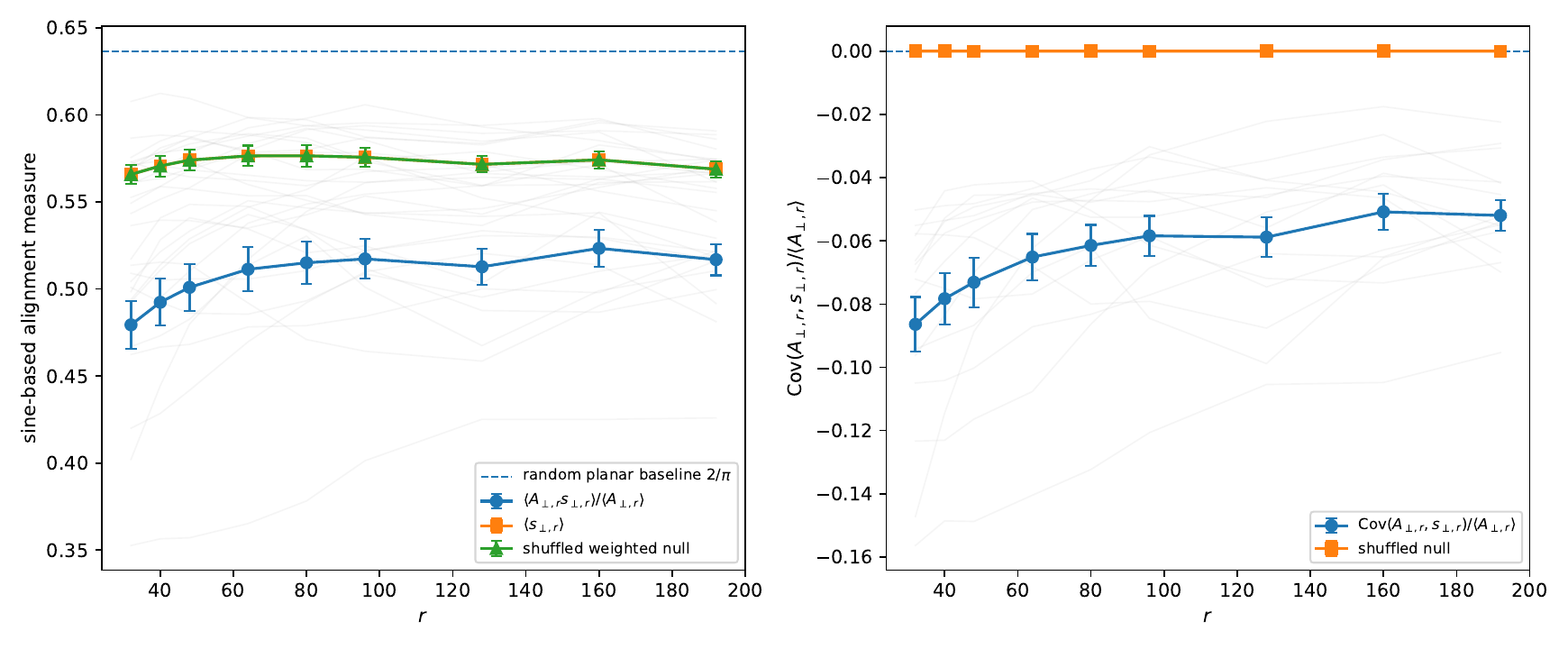}
\caption{\footnotesize Weighted and unweighted means of $s_r$ in fifteen
$320^3$ full MHD JHTDB subvolumes. Left: $\langle s_r\rangle_A$,
$\langle s_r\rangle$, and a shuffled-weight check. Right: the normalized
amplitude--angle covariance and its shuffled control. Faint curves show
individual subvolumes and error bars are standard errors across subvolumes.}
\label{fig_app_weighted_covariance}
\end{figure*}

\begin{figure*}[t]
\centering
\includegraphics[width=\columnwidth]{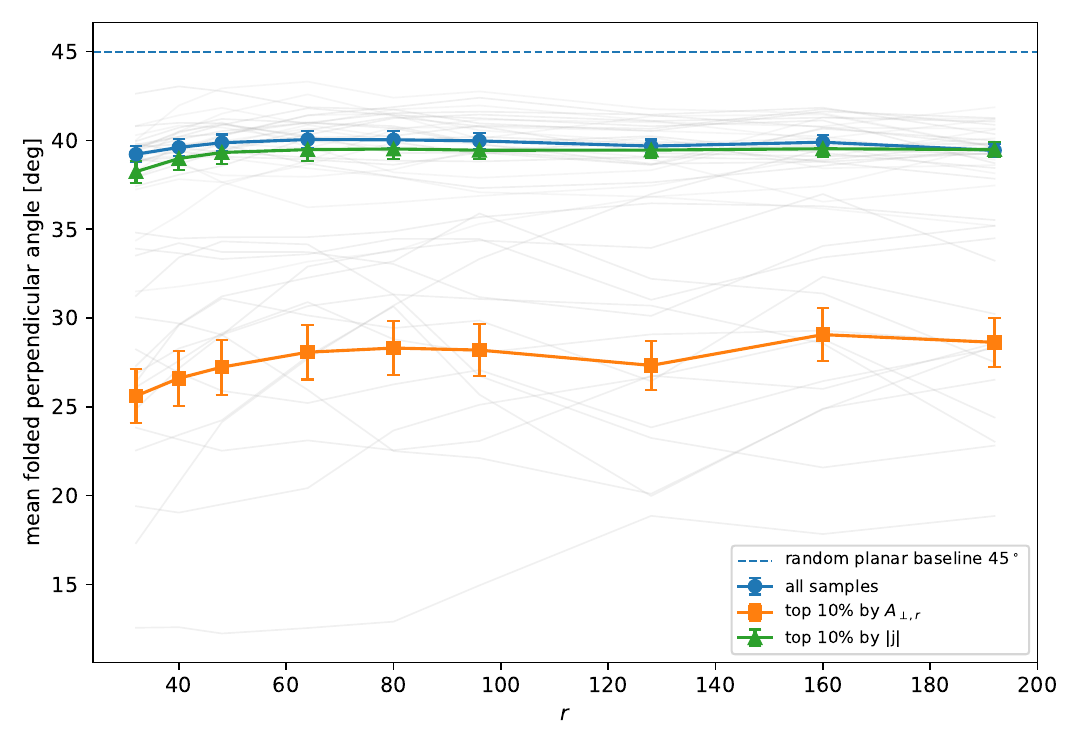}
\caption{\footnotesize Folded mean angle between Els\'asser increments for the full sample, the top $10\%$ by $A_r$, and the top
\(10\%\) by current density magnitude in the fifteen JHTDB subvolumes.}
\label{fig_app_current_densityweighting_Vishniac}
\end{figure*}

\subsection{Low order structure function}

To test the scale dependence of the amplitude distribution (suggested by G. Eyink, private communicatin), we use the normalized moments $I_p$ defined by Equation (\ref{Eyink1}). This quantity is constant for a self-similar amplitude distribution whose shape is independent of scale. Its growth toward smaller separation means that the
$p$th moment is becoming increasingly concentrated in the large-amplitude
tail \citep{Sreenivasanetal1996,Caoetal1996,Chenetal2005}. Previous conditional and intermittent studies anticipated this behavior
\citep{MalletEtAl2016,MalletSchekochihin2017}.

For a positive $X_r\in \{ |\delta_r\boldsymbol z^+_\perp|, |\delta_r\boldsymbol z^-_\perp|,  A_r^{1/2}\}$;
\begin{equation}
 \left.\frac{\partial}{\partial p}\ln\langle X_r^p\rangle\right|_{p=0}
 =\langle\ln X_r\rangle.
 \label{eq-app_low_order}
\end{equation}
We fit both $\langle\ln X_r\rangle$ and $\tfrac12\ln\langle X_r^2\rangle$ as functions of $\ln r$; see Table \ref{tabel_appen_low_orders}.

\begin{table}[t]
\centering
\caption{\footnotesize Amplitude slopes in the $448^3$ JHTDB subvolume. The first slope results from $\langle\ln X_r\rangle$ and the second from $\tfrac12\ln\langle X_r^2\rangle$. Fits use $32\leq r\leq160$.}
\label{tabel_appen_low_orders}
{\footnotesize
\setlength{\tabcolsep}{2.5pt}
\begin{tabular}{cccc}
\hline\hline
Amplitude & \shortstack{Slope of\\$\langle\ln X_r\rangle$} & \shortstack{Slope of\\$\tfrac12\ln\langle X_r^2\rangle$} & Difference\\
\hline
$|\delta_r\boldsymbol z^+_\perp|$ & $0.315\pm0.009$ & $0.242\pm0.008$ & 0.073\\
$|\delta_r\boldsymbol z^-_\perp|$ & $0.351\pm0.007$ & $0.289\pm0.012$ & 0.062\\
$A_r^{1/2}$       & $0.333\pm0.008$ & $0.267\pm0.010$ & 0.066\\
\hline
\end{tabular}}
\end{table}

\subsection{Fourier randomization tests}

Following the suggestion by A. Beresnyak (private communications), we also measure the amplitude--angle covariance in two Fourier-randomized fields made from three stored full-MHD fields. In the tensor-phase field, all components of a given Fourier mode receive the same random phase. This preserves the spectral tensor of each mode and the velocity--magnetic geometry at the same wavevector, while removing the phase relations between different modes. The shell-isotropized Gaussian field preserves only the shell-averaged Els\"asser spectra. Figure~\ref{figur:appfouriernullBeresnyak} shows that the shell-isotropized field has a negligible covariance, while the tensor-phase field contains only part of the covariance measured in the MHD data.

\begin{figure*}[t]
\centering
\includegraphics[width=\columnwidth]{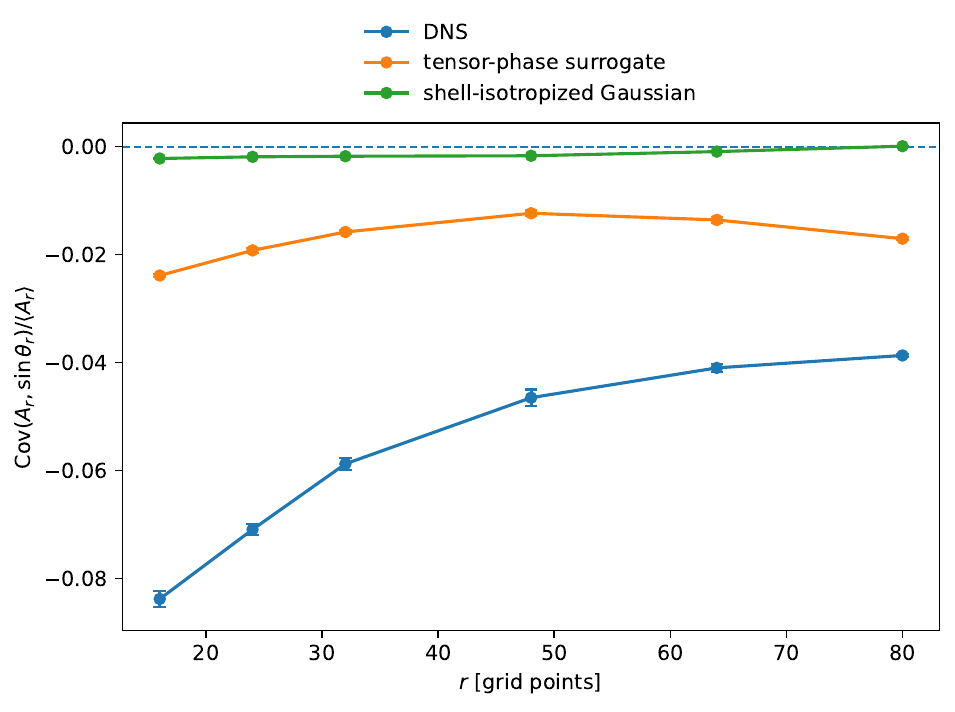}
\caption{\footnotesize Normalized amplitude-angle covariance in JHTDB fields and in two Fourier-randomized fields. The shell-isotropized Gaussian field gives values close to zero while the tensor-phase field gives an intermediate covariance.}
\label{figur:appfouriernullBeresnyak}
\end{figure*}

\subsection{Wind observations}
\label{app:windsubsection}

Magnetic-field measurements from the Wind MFI instrument and proton density and velocity from the Wind 3DP instrument are placed on a common 30-s cadence \citep{NASA1,NASA2,NASA3,Kovaletal2021,Wind3DPMoments}. The primary sample contains fifty 24-hour intervals. The Wind Elsässer fields are
$$
 {\boldsymbol z}^\pm(t)={\boldsymbol v}(t)
 \pm\frac{{\boldsymbol B}(t)}{\sqrt{\mu_0m_p n_p(t)}}.
$$
The time-lagged increments, Els\"asser amplitude, and angle are
$$ \delta_\tau\boldsymbol z^\pm(t)=\boldsymbol z^\pm(t+\tau)-\boldsymbol z^\pm(t),\;
 A_\tau=|\delta_\tau\boldsymbol z^+|\,|\delta_\tau\boldsymbol z^-|,$$
 $$\theta_\tau=\arccos \frac{|\delta_\tau\boldsymbol z^+\!\cdot\delta_\tau\boldsymbol z^-|}{|\delta_\tau\boldsymbol z^+|\,|\delta_\tau\boldsymbol z^-|},
 \;s_\tau=\sin\theta_\tau.
$$

We use the full three-component increments at time separations $\tau=60,120,240,480,960$, and $1920$ s. Under Taylor sampling, these are one-dimensional measurements along the spacecraft path rather than the local-perpendicular increments used in the simulations \citep{Taylor1938}. Figure~\ref{fig:append_wind-data} shows the measured angles and covariance; shuffling the amplitudes removes the covariance. For $X_\tau=|\delta_\tau\boldsymbol z^\pm|$ over $240\leq\tau\leq15360$ s, the slope obtained from $\langle\ln X_\tau\rangle$ exceeds the slope obtained from $\tfrac12\ln\langle X_\tau^2\rangle$ by $0.096\pm0.005$.

\begin{figure*}[t]
\centering
\includegraphics[width=\textwidth]{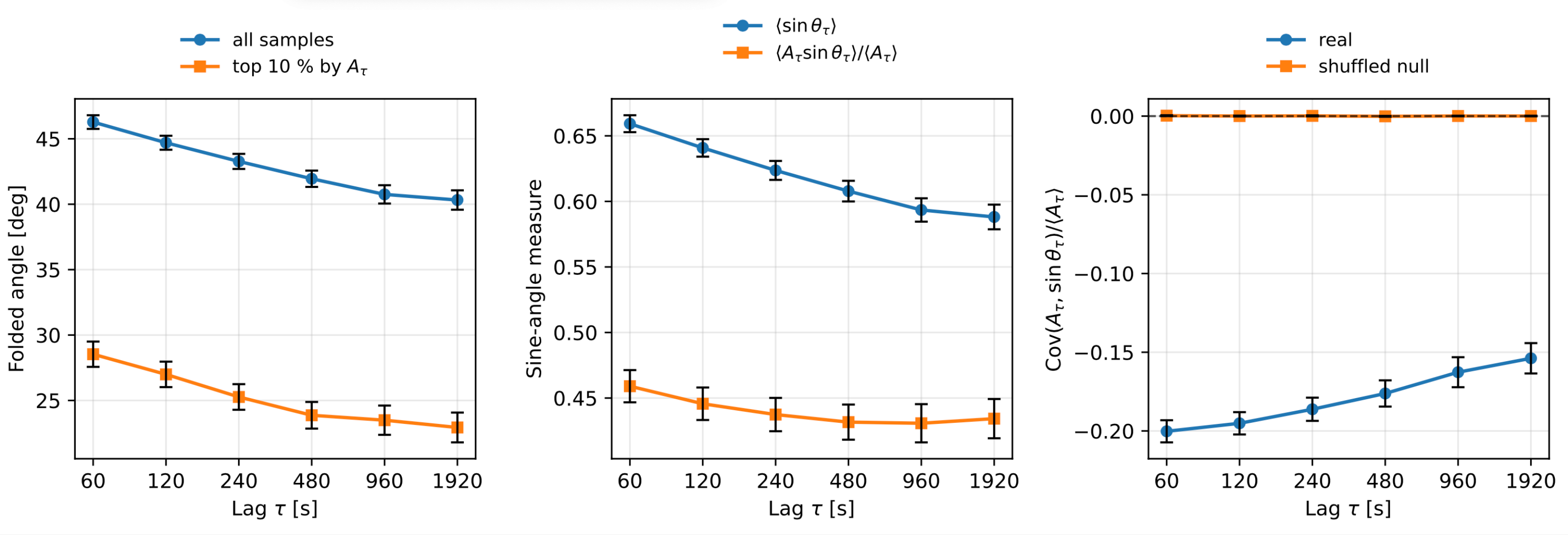}
\caption{\footnotesize Wind results from fifty 24-hour intervals. Left: folded angle for all increments and for the top 10\% ranked by the Els\"asser amplitude $A_\tau$. Middle: unweighted and amplitude-weighted means of $s_\tau$. Right: normalized amplitude--angle covariance and the result after shuffling the amplitudes. Error bars are standard errors across intervals.}
\label{fig:append_wind-data}
\end{figure*}

}

\bibliographystyle{aasjournalv7}
\bibliography{NonDynamicAlignment}

@article{MullerGrappin2005,
  title = {Spectral Energy Dynamics in Magnetohydrodynamic Turbulence},
  author = {M{\"u}ller, Wolf-Christian and Grappin, Roland},
  journal = {Phys. Rev. Lett.},
  volume = {95},
  issue = {11},
  pages = {114502},
  numpages = {4},
  year = {2005},
  month = {Sep},
  publisher = {American Physical Society},
  doi = {10.1103/PhysRevLett.95.114502},
  url = {https://link.aps.org/doi/10.1103/PhysRevLett.95.114502}
}

@article{Boldyrev2005,
doi = {10.1086/431649},
url = {https://doi.org/10.1086/431649},
year = {2005},
month = {may},
publisher = {},
volume = {626},
number = {1},
pages = {L37},
author = {Boldyrev, Stanislav},
title = {On the Spectrum of Magnetohydrodynamic Turbulence},
journal = {The Astrophysical Journal}
}

@article{Kraichnan1965,
    author = {Kraichnan, Robert H.},
    title = {Inertial Range Spectrum of Hydromagnetic Turbulence},
    journal = {The Physics of Fluids},
    volume = {8},
    number = {7},
    pages = {1385-1387},
    year = {1965},
    month = {07},
    issn = {0031-9171},
    doi = {10.1063/1.1761412},
    url = {https://doi.org/10.1063/1.1761412},
    eprint = {https://pubs.aip.org/aip/pfl/article-pdf/8/7/1385/12617184/1385_1_online.pdf},
}

@misc{Sioulas2024,
      title={Scale-Dependent Dynamic Alignment in MHD Turbulence: Insights into Intermittency, Compressibility, and Imbalance Effects}, 
      author={Nikos Sioulas and Marco Velli and Alfred Mallet and Trevor A. Bowen and B. D. G. Chandran and Chen Shi and S. S. Cerri and Ioannis Liodis and Tamar Ervin and Davin E. Larson},
      year={2024},
      eprint={2407.03649},
      archivePrefix={arXiv},
      primaryClass={physics.space-ph},
      url={https://arxiv.org/abs/2407.03649}, 
}

@article{Wicks_2013,
   title={Alignment and Scaling of Large-Scale Fluctuations in the Solar Wind},
   volume={110},
   ISSN={1079-7114},
   url={http://dx.doi.org/10.1103/PhysRevLett.110.025003},
   DOI={10.1103/physrevlett.110.025003},
   number={2},
   journal={Physical Review Letters},
   publisher={American Physical Society (APS)},
   author={Wicks, R. T. and Mallet, A. and Horbury, T. S. and Chen, C. H. K. and Schekochihin, A. A. and Mitchell, J. J.},
   year={2013},
   month=Jan }

@ARTICLE{Iroshinkov1964,
       author = {{Iroshnikov}, P.~S.},
        title = "{Turbulence of a Conducting Fluid in a Strong Magnetic Field}",
      journal = {\sovast},
         year = 1964,
        month = feb,
       volume = {7},
        pages = {566},
       adsurl = {https://ui.adsabs.harvard.edu/abs/1964SvA.....7..566I}
}

@article{MaronGoldreich2001,
doi = {10.1086/321413},
url = {https://doi.org/10.1086/321413},
year = {2001},
month = {jun},
publisher = {},
volume = {554},
number = {2},
pages = {1175},
author = {Maron, Jason and Goldreich, Peter},
title = {Simulations of Incompressible Magnetohydrodynamic Turbulence},
journal = {The Astrophysical Journal}
}

@article{BiskampMuller2000,
    author = {Biskamp, Dieter and Müller, Wolf-Christian},
    title = {Scaling properties of three-dimensional isotropic magnetohydrodynamic turbulence},
    journal = {Physics of Plasmas},
    volume = {7},
    number = {12},
    pages = {4889-4900},
    year = {2000},
    month = {12},
    issn = {1070-664X},
    doi = {10.1063/1.1322562},
    url = {https://doi.org/10.1063/1.1322562},
    eprint = {https://pubs.aip.org/aip/pop/article-pdf/7/12/4889/19071562/4889_1_online.pdf},
}

@article{ChoLazarian2003,
  author   = {Cho, Jungyeon and Lazarian, A.},
  title    = {Compressible magnetohydrodynamic turbulence: mode coupling, scaling relations, anisotropy, viscosity-damped regime and astrophysical implications},
  journal  = {Monthly Notices of the Royal Astronomical Society},
  volume   = {345},
  number   = {1},
  pages    = {325--339},
  year     = {2003},
  doi      = {10.1046/j.1365-8711.2003.06941.x},
  issn     = {0035-8711}
}

@article{ChoLazarianVishniac2002,
doi = {10.1086/324186},
url = {https://doi.org/10.1086/324186},
year = {2002},
month = {jan},
publisher = {},
volume = {564},
number = {1},
pages = {291},
author = {Cho, Jungyeon and Lazarian, Alex and Vishniac, Ethan T.},
title = {Simulations of Magnetohydrodynamic Turbulence in a Strongly Magnetized Medium},
journal = {The Astrophysical Journal}
}

@article{Matthaeus2012,
doi = {10.1088/0004-637X/750/2/103},
url = {https://doi.org/10.1088/0004-637X/750/2/103},
year = {2012},
month = {apr},
publisher = {The American Astronomical Society},
volume = {750},
number = {2},
pages = {103},
author = {Matthaeus, W. H. and Servidio, S. and Dmitruk, P. and Carbone, V. and Oughton, S. and Wan, M. and Osman, K. T.},
title = {LOCAL ANISOTROPY, HIGHER ORDER STATISTICS, AND TURBULENCE SPECTRA},
journal = {The Astrophysical Journal}
}

@misc{JHTDBMHD,
  author = {Aluie, Hussein and Eyink, Gregory L. and Vishniac, Ethan T. and others},
  title = {Forced MHD Turbulence Dataset},
  year = {2012},
  howpublished = {Johns Hopkins Turbulence Databases},
  doi = {10.7281/T1930RBS}
}

@article{Beresnyak2011,
  author = {Beresnyak, Andrey},
  year = {2011},
  journal = {Physical Review Letters},
  volume = {106},
  pages = {075001},
  doi = {10.1103/PhysRevLett.106.075001}
}

@article{Beresnyak2012,
  author = {Beresnyak, Andrey},
  year = {2012},
  journal = {Monthly Notices of the Royal Astronomical Society},
  volume = {422},
  pages = {3495},
  doi = {10.1111/j.1365-2966.2012.20859.x}
}

@article{Beresnyak2014,
  author = {Beresnyak, Andrey},
  year = {2014},
  journal = {The Astrophysical Journal Letters},
  volume = {784},
  pages = {L20},
  doi = {10.1088/2041-8205/784/2/L20}
}

@article{Beresnyak2015,
  author = {Beresnyak, Andrey},
  year = {2015},
  journal = {The Astrophysical Journal Letters},
  volume = {801},
  pages = {L9},
  doi = {10.1088/2041-8205/801/1/L9}
}

@article{BeresnyakLazarian2009,
doi = {10.1088/0004-637X/702/2/1190},
url = {https://doi.org/10.1088/0004-637X/702/2/1190},
year = {2009},
month = {aug},
publisher = {The American Astronomical Society},
volume = {702},
number = {2},
pages = {1190},
author = {Beresnyak, A. and Lazarian, A.},
title = {COMPARISON OF SPECTRAL SLOPES OF MAGNETOHYDRODYNAMIC AND HYDRODYNAMIC TURBULENCE AND MEASUREMENTS OF ALIGNMENT EFFECTS},
journal = {The Astrophysical Journal}
}

@article{BeresnyakLazarian2006,
  author = {Beresnyak, Andrey and Lazarian, Alex},
  year = {2006},
  journal = {The Astrophysical Journal Letters},
  volume = {640},
  pages = {L175},
  doi = {10.1086/503708}
}

@article{Boldyrev2006,
  author = {Boldyrev, Stanislav},
  year = {2006},
  journal = {Physical Review Letters},
  volume = {96},
  pages = {115002},
  doi = {10.1103/PhysRevLett.96.115002}
}

@article{Caoetal1996,
  author = {Cao, Ning and Chen, Shiyi and Sreenivasan, Katepalli R.},
  year = {1996},
  journal = {Physical Review Letters},
  volume = {77},
  pages = {3799},
  doi = {10.1103/PhysRevLett.77.3799}
}

@article{ChandranSchekochihinMallet2015,
  author = {Chandran, Benjamin D. G. and Schekochihin, Alexander A. and Mallet, Alfred},
  year = {2015},
  journal = {The Astrophysical Journal},
  volume = {807},
  pages = {39},
  doi = {10.1088/0004-637X/807/1/39}
}

@article{Chenetal2005,
  author = {Chen, Shiyi and Dhruva, Bhaskar and Kurien, Susan and Sreenivasan, Katepalli R. and Taylor, Michael A.},
  year = {2005},
  journal = {Journal of Fluid Mechanics},
  volume = {533},
  pages = {183},
  doi = {10.1017/S002211200500443X}
}

@article{Eyinketal2013,
  author = {Eyink, Gregory and Vishniac, Ethan and Lalescu, Cristian and others},
  year = {2013},
  journal = {Nature},
  volume = {497},
  pages = {466},
  doi = {10.1038/nature12128}
}

@article{Frank2012Price,
  author = {Frank, Steven A.},
  year = {2012},
  journal = {Journal of Evolutionary Biology},
  volume = {25},
  pages = {1002},
  doi = {10.1111/j.1420-9101.2012.02498.x}
}

@article{Galtier2011,
  author = {Galtier, Sebastien},
  year = {2011},
  journal = {Comptes Rendus Physique},
  volume = {12},
  pages = {151},
  doi = {10.1016/j.crhy.2010.11.006}
}

@misc{Kovaletal2021,
  author = {Koval, Andriy and Lepping, Ronald P. and Szabo, Adam},
  title = {Wind Magnetic Field Investigation (MFI) Composite Data},
  year = {2021},
  howpublished = {Space Physics Data Facility},
  doi = {10.48322/AV38-WN55}
}

@article{JHTB1,
  author = {Li, Yi and Perlman, Eric and Wan, Minping and others},
  year = {2008},
  journal = {Journal of Turbulence},
  volume = {9},
  pages = {N31},
  doi = {10.1080/14685240802376389}
}

@misc{Wind3DPMoments,
  author = {Lin, Robert P. and Bale, Stuart D. and Wilson, III, Lynn B.},
  title = {Wind 3DP PESA Moments Data},
  year = {2021},
  howpublished = {Space Physics Data Facility and University of California, Berkeley},
  doi = {10.48322/s8e6-aw08}
}

@article{NASA3,
  author = {Lin, Robert P. and Anderson, Kenneth A. and Ashford, Steven and others},
  year = {1995},
  journal = {Space Science Reviews},
  volume = {71},
  pages = {125},
  doi = {10.1007/BF00751328}
}

@article{NASA2,
  author = {Lepping, Ronald P. and Acuna, Mario H. and Burlaga, Leonard F. and others},
  year = {1995},
  journal = {Space Science Reviews},
  volume = {71},
  pages = {207},
  doi = {10.1007/BF00751330}
}

@article{MalletSchekochihin2017,
  author = {Mallet, Alfred and Schekochihin, Alexander A.},
  year = {2017},
  journal = {Monthly Notices of the Royal Astronomical Society},
  volume = {466},
  pages = {3918},
  doi = {10.1093/mnras/stw3251}
}

@article{Podesta2009,
author = {Podesta, J. J. and Chandran, B. D. G. and Bhattacharjee, A. and Roberts, D. A. and Goldstein, M. L.},
title = {Scale-dependent angle of alignment between velocity and magnetic field fluctuations in solar wind turbulence},
journal = {Journal of Geophysical Research: Space Physics},
volume = {114},
number = {A1},
pages = {},
doi = {https://doi.org/10.1029/2008JA013504},
url = {https://agupubs.onlinelibrary.wiley.com/doi/abs/10.1029/2008JA013504},
eprint = {https://agupubs.onlinelibrary.wiley.com/doi/pdf/10.1029/2008JA013504},
year = {2009}
}

@article{MalletSchekochihinChandran2015,
  author = {Mallet, Alfred and Schekochihin, Alexander A. and Chandran, Benjamin D. G.},
  year = {2015},
  journal = {Monthly Notices of the Royal Astronomical Society: Letters},
  volume = {449},
  pages = {L77},
  doi = {10.1093/mnrasl/slv021}
}

@article{MalletEtAl2016,
  author = {Mallet, Alfred and Schekochihin, Alexander A. and Chandran, Benjamin D. G. and others},
  year = {2016},
  journal = {Monthly Notices of the Royal Astronomical Society},
  volume = {459},
  pages = {2130},
  doi = {10.1093/mnras/stw802}
}

@article{Mason2006,
  author = {Mason, Joanne and Cattaneo, Fausto and Boldyrev, Stanislav},
  year = {2006},
  journal = {Physical Review Letters},
  volume = {97},
  pages = {255002},
  doi = {10.1103/PhysRevLett.97.255002}
}

@misc{NASA1,
  author = {{NASA Space Physics Data Facility}},
  title = {Coordinated Data Analysis Web (CDAWeb)},
  year = {2026},
  howpublished = {NASA Goddard Space Flight Center},
  url = {https://cdaweb.gsfc.nasa.gov/}
}

@article{PerezEtAl2012,
  author = {Perez, Jean Carlos and Mason, Joanne and Boldyrev, Stanislav and Cattaneo, Fausto},
  year = {2012},
  journal = {Physical Review X},
  volume = {2},
  pages = {041005},
  doi = {10.1103/PhysRevX.2.041005}
}

@inproceedings{JHTB2,
  author = {Perlman, Eric and Burns, Randal and Li, Yi and Meneveau, Charles},
  year = {2007},
  booktitle = {Proceedings of the 2007 ACM/IEEE Conference on Supercomputing},
  publisher = {Association for Computing Machinery},
  address = {New York, NY},
  doi = {10.1145/1362622.1362654}
}

@article{PolitanoPouquet1998,
  author = {Politano, Helene and Pouquet, Annick},
  year = {1998},
  journal = {Physical Review E},
  volume = {57},
  pages = {R21},
  doi = {10.1103/PhysRevE.57.R21}
}

@article{Price1970,
  author = {Price, George R.},
  year = {1970},
  journal = {Nature},
  volume = {227},
  pages = {520},
  doi = {10.1038/227520a0}
}

@article{Price1972,
  author = {Price, George R.},
  year = {1972},
  journal = {Annals of Human Genetics},
  volume = {35},
  pages = {485},
  doi = {10.1111/j.1469-1809.1957.tb01874.x}
}

@article{Schekochihin2022,
  author = {Schekochihin, Alexander A.},
  year = {2022},
  journal = {Journal of Plasma Physics},
  volume = {88},
  pages = {155880501},
  doi = {10.1017/S0022377822000721}
}

@misc{SioulasEtAl2024,
  author = {Sioulas, Nikolaos and Velli, Marco and Mallet, Alfred and others},
  title = {Scale-Dependent Dynamic Alignment in MHD Turbulence: Insights into Intermittency, Compressibility, and Imbalance Effects},
  year = {2024},
  eprint = {2407.03649},
  archiveprefix = {arXiv},
  doi = {10.48550/arXiv.2407.03649}
}

@article{Sreenivasanetal1996,
  author = {Sreenivasan, Katepalli R. and Vainshtein, Samuel I. and Bhiladvala, Rustom and others},
  year = {1996},
  journal = {Physical Review Letters},
  volume = {77},
  pages = {1488},
  doi = {10.1103/PhysRevLett.77.1488}
}

@article{Taylor1938,
  author = {Taylor, Geoffrey I.},
  year = {1938},
  journal = {Proceedings of the Royal Society of London. Series A},
  volume = {164},
  pages = {476},
  doi = {10.1098/rspa.1938.0032}
}

@article{GoldreichSridhar1995,
  author = {Goldreich, Peter and Sridhar, S.},
  title = {Toward a Theory of Interstellar Turbulence. II. Strong Alfv\'enic Turbulence},
  year = {1995},
  journal = {The Astrophysical Journal},
  volume = {438},
  pages = {763},
  doi = {10.1086/175121}
}

@article{Kolmogorov1941,
  author  = {Kolmogorov, A. N.},
  title   = {The Local Structure of Turbulence in Incompressible Viscous Fluid for Very Large Reynolds Numbers},
  journal = {Doklady Akademii Nauk SSSR},
  year    = {1941},
  volume  = {30},
  pages   = {299--303}
}

@article{Onsager1949,
  author  = {Onsager, L.},
  title   = {Statistical Hydrodynamics},
  journal = {Nuovo Cimento Supplemento},
  year    = {1949},
  volume  = {6},
  pages   = {279--287},
  doi     = {10.1007/BF02780991}
}

@article{CaflischKlapperSteele1997,
  author  = {Caflisch, R. E. and Klapper, I. and Steele, G.},
  title   = {Remarks on Singularities, Dimension and Energy Dissipation for Ideal Hydrodynamics and MHD},
  journal = {Communications in Mathematical Physics},
  year    = {1997},
  volume  = {184},
  pages   = {443--455},
  doi     = {10.1007/s002200050067}
}

@article{Obukhov1962,
  author  = {Obukhov, A. M.},
  title   = {Some Specific Features of Atmospheric Turbulence},
  journal = {Journal of Fluid Mechanics},
  year    = {1962},
  volume  = {13},
  pages   = {77--81},
  doi     = {10.1017/S0022112062000506}
}

@article{Kolmogorov1962,
  author  = {Kolmogorov, A. N.},
  title   = {A Refinement of Previous Hypotheses Concerning the Local Structure of Turbulence in a Viscous Incompressible Fluid at High Reynolds Number},
  journal = {Journal of Fluid Mechanics},
  year    = {1962},
  volume  = {13},
  pages   = {82--85},
  doi     = {10.1017/S0022112062000518}
}

@incollection{ParisiFrisch1985,
  author    = {Frisch, U. and Parisi, G.},
  title     = {On the Singularity Structure of Fully Developed Turbulence},
  booktitle = {Turbulence and Predictability in Geophysical Fluid Dynamics and Climate Dynamics},
  editor    = {Ghil, M. and Benzi, R. and Parisi, G.},
  publisher = {North-Holland},
  address   = {Amsterdam},
  year      = {1985},
  pages     = {84--87}
}

@book{Frisch1995,
  author    = {Frisch, U.},
  title     = {Turbulence: The Legacy of A. N. Kolmogorov},
  publisher = {Cambridge University Press},
  address   = {Cambridge},
  year      = {1995}
}

@article{Carbone1993,
  author  = {Carbone, V.},
  title   = {Cascade Model for Intermittency in Fully Developed Magnetohydrodynamic Turbulence},
  journal = {Physical Review Letters},
  year    = {1993},
  volume  = {71},
  pages   = {1546--1549},
  doi     = {10.1103/PhysRevLett.71.1546}
}

@article{PolitanoPouquet1995,
  author  = {Politano, H. and Pouquet, A.},
  title   = {Model of Intermittency in Magnetohydrodynamic Turbulence},
  journal = {Physical Review E},
  year    = {1995},
  volume  = {52},
  pages   = {636--641},
  doi     = {10.1103/PhysRevE.52.636}
}

@article{MullerBiskamp2000,
  author  = {Muller, W.-C. and Biskamp, D.},
  title   = {Scaling Properties of Three-Dimensional Magnetohydrodynamic Turbulence},
  journal = {Physical Review Letters},
  year    = {2000},
  volume  = {84},
  pages   = {475--478},
  doi     = {10.1103/PhysRevLett.84.475}
}

@article{JiangEtAl2026,
  author  = {Jiang, B. and Yang, Y. and Ye, Y. and Li, C. and Wang, J. and Oughton, S. and Matthaeus, W. H. and Wan, M.},
  title   = {Assessment of Refined Similarity Hypothesis in Magnetohydrodynamic Turbulence},
  journal = {Journal of Fluid Mechanics},
  year    = {2026},
  volume  = {1037},
  pages   = {A37},
  doi     = {10.1017/jfm.2026.11686}
}

\end{document}